%% file: SplitReg_arXiv.tex
\documentclass[12pt]{article}%
\usepackage{amssymb}
\usepackage{amsfonts}
\usepackage{graphicx, epsfig}
\usepackage{amsmath}
\usepackage{amsthm}
\usepackage{morefloats}
\usepackage{placeins}
\usepackage{rotating}
\usepackage{natbib}
\usepackage{url} 
\usepackage{caption,subcaption}
\usepackage{booktabs}
\usepackage{multirow}
\usepackage{afterpage}
\providecommand{\U}[1]{\protect\rule{.1in}{.1in}}

\usepackage{natbib}
\setcitestyle{aysep={}} 
\newtheorem{theorem}{Theorem}

\newtheorem{prop}{Proposition}
\newcommand{\bbet}{\boldsymbol{\beta}}
\newcommand{\hbbet}{\hat{\boldsymbol{\beta}}}
\newcommand{\hbeta}{\hat{\beta}}
\DeclareMathOperator{\Tr}{Tr}
\DeclareMathOperator{\sign}{sign}
\makeatletter
\def\p@subsection{}
\makeatother

\usepackage{placeins}
\usepackage{booktabs}
\usepackage{multirow}
\usepackage{rotating}
\usepackage{afterpage}

\begin{document}
	
	\title{\textbf{Split Regularized Regression }}
	\author{Anthony-Alexander Christidis\\Department of Statistics, University of British Columbia\\(anthony.christidis@stat.ubc.ca)  \\Laks Lakshmanan \\Department of Computer Science, University of British Columbia\\(laks@cs.ubc.ca)\\Ezequiel Smucler \\Department of Mathematics and Statistics, Universidad Torcuato Di Tella\\(esmucler@utdt.edu)\\ and \\Ruben Zamar\\Department of Statistics, University of British Columbia\\(ruben@stat.ubc.ca)}
	\date{}
	\maketitle
	
	\begin{abstract}
		{\normalsize We propose an approach for 
			fitting linear regression models that splits the set of covariates into groups. The optimal split of the variables into groups and the regularized estimation of the regression coefficients are performed by minimizing  an objective function that encourages sparsity within each group and diversity among them. 
			The estimated coefficients are then pooled together to form the final fit. Our procedure works on top of a given penalized linear regression estimator (e.g., Lasso, elastic net) by fitting it to possibly overlapping groups of features, encouraging diversity among these groups to reduce the correlation of the corresponding predictions. For the case of two groups, elastic net penalty and orthogonal predictors, we give a closed form solution for the regression coefficients in each group. We establish the consistency of our method with the number of predictors possibly  increasing with the sample size. An extensive simulation study and real-data applications show that in general the proposed method improves the prediction accuracy of the base 
			estimator used in the procedure. Possible extensions to GLMs and other models are discussed. The supplemental material for this article, available online, contains the proofs of our theoretical results and the full results of our simulation study.
		}
		
	\end{abstract}
	
	\bibliographystyle{natbib}

	\renewcommand{\baselinestretch}{1.45}

	{\normalsize \bigskip}
	
	
	{\normalsize \newpage\renewcommand{\baselinestretch}{1.5}{\normalsize
		}
		
		\renewcommand{\baselinestretch}{1.45}
		
		\section{Introduction} \label{sec:Introduction}
		Suppose we have training data $\left( \mathbf{y}, \mathbf{X}\right)$, where $\mathbf{y}\in \mathbb{R}^{n}$ is a  vector of response variables
		and $\mathbf{X}\in \mathbb{R}^{n\times p}$ is a design  matrix comprising  $n$  measurements  on $p$
		features.  We consider the linear regression model  $ \mathbf{y} = \mathbf{X}
		\boldsymbol{\beta }+\boldsymbol{\varepsilon}$, where $\boldsymbol{\beta}\in \mathbb{R}^{p}$  is the vector of regression coefficients and $\boldsymbol{\varepsilon} \in \mathbb{R}^{n}$ is a vector of independent regression errors.
		We assume that the response $\mathbf{y}^{\prime }=(y_{1},\dots ,y_{n})$,   and the entries of the design matrix,  $x_{i,j}$,
		$i=1,\dots ,n$ and $j=1,...,p$,
		are standardized so that 
		\begin{align*}
		\frac{1}{n}\sum\limits_{i=1}^{n}x_{i,j}=0,\quad 
		\frac{1}{n}\sum\limits_{i=1}^{n}x_{i,j}^{2}=1,\quad 1\leq j \leq p, \quad 
		\frac{1}{n}\sum\limits_{i=1}^{n}y_{i}=0, \quad 
		\frac{1}{n}\sum\limits_{i=1}^{n}y_{i}^{2}=1.
		\end{align*}
		The situations we are mainly interested on are those in which the number of variables  $p$ is  large, of the same order or larger than $n$, because in these cases  there is a good chance for a favorable  
		bias-variance trade-off.
		
		Regularization  can  improve estimation and prediction in linear regression models, allowing  their  application in situations where the number of explanatory variables is  larger than the number of observations. In this setting variance reduction is achieved by shrinking  the absolute size of the estimators.  Ridge regression \citep{Ridge}, non-negative garrote \citep{Garrotte}, Lasso \citep{Lasso}, elastic net \citep{EN} and other similar methods proceed in this general way. When some coefficients are fully shrunk to zero, as  in the case of Lasso, regularization also provides an effective way for variable selection. 
		
		A key question  in the context of regularization is: which coefficients should be shrunk and  by how much?   In general, this  issue is decided by minimizing a penalized least squares objective function of the form 
		\begin{equation}
		O(\mathbf{y},\mathbf{X},\boldsymbol{\beta })= \frac{1}{2n}\Vert \mathbf{y}-\mathbf{X}
		\boldsymbol{\beta }\Vert ^{2}_2+ \lambda P(\boldsymbol{\beta }),
		\label{eq:lasso}
		\end{equation}
		where $ \Vert \mathbf{a}\Vert_2$ is the Euclidean norm of $\mathbf{a}$, 
		$ P(\boldsymbol{\beta })$ is a penalty function that encourages shrinkage, e.g.  $P(\boldsymbol{\beta })=\sum^p_{j=1} \beta_j^2$ in the case of ridge regression and $P(\boldsymbol{\beta })=\sum^p_{j=1} |\beta_j|$ in the case of Lasso.  The constant $\lambda\geq0$  determines the amount of shrinkage. 
		
		On a theoretical level, \cite{christidis:2018} shows that variance reduction can  be   achieved by splitting the set of explanatory variables into subsets and applying linear regression to each subset separately. They  show that  an appropriate mix  of splitting and shrinkage could  be very effective  for achieving a favorable variance-bias trade-off. But they do not propose a procedure for exploiting this  opportunity in practice. 
		In general, one could exhaustively search over all possible groupings of the
		variables into different models and choose the one with the lowest estimated
		prediction error (e.g. using cross-validation), but this is computationally
		unfeasible. Let $h_i(p_{1},p_{2},...,p_{G})$ be the number of elements in the sequence $p_{1},p_{2},...,p_{G}$ that are equal to $ i $, $ i=1,\dots,\big\lfloor\frac{p-(G-2)}{2}\big\rfloor$. The number of possible splits of $p$ features into $G$ 
		groups of sizes $p_g \in \mathbb{N}^+$, $g=1,...,G$ is 
		\begin{equation*}
		\sum_{p_{1}\leq p_{2}\leq\cdots\leq p_{G}}\Bigg[\frac{p!}{p_{1}!p_{2}! \dots p_{G}!} \prod_{i=1}^{\big\lfloor\frac{p-(G-2)}{2}\big\rfloor}\frac{1}{h_i(p_{1},p_{2},...,p_{G})!}\Bigg]
		\end{equation*}
		For example, the number of possible splits of $ p=15 $ features into $ G=3 $ groups is 2,375,101. The number of possible splits is much larger if we allow the variables to be shared by the different groups. 
		
		Therefore, important questions arising  in this context are: {\em is there a practical way for finding the best (or at least  a good) split of  the variables and at the same time which coefficients should be shrunk and  by how much? } In Section \ref{sec:TheMethod} we introduce a procedure  called SplitReg that addresses these questions and study its properties  in some simple but illustrative cases. The computational complexity of SplitReg is approximately equal to that of the  base regularized estimator multiplied by the number of groups.

		The rest of this article is organized as follows.   
		In Section 
		\ref{sec:alg} we propose an algorithm to compute the proposed estimators, to 
		choose their tuning parameters and to aggregate the predictions
		from the constructed models. We prove the consistency of our proposal    in Section \ref{sec:cons}. In Section \ref{sec:sim} we conduct an extensive  simulation study  to compare
		the performance of SplitReg, with regards to
		prediction accuracy,  against  several
		state of the art alternatives. We apply all the procedures considered in the simulation study to real 
		data-sets in Section \ref{sec:real}. Finally, some conclusions and
		possible extensions are discussed in Section \ref{sec:disc}. 
		Technical proofs and additional simulation results
		are provided in the supplementary material for this article.
		
		\section{Split-Regularized Regression (SplitReg)} \label{sec:TheMethod}

		We will show here that an approach for addressing the questions posed at the end of Section \ref{sec:Introduction}  is to find a minimizer $\hbbet =\left( \hbbet^1,\cdots,\hbbet^G \right)$ of an objective function  of the form  
		\begin{equation}
		O(\mathbf{y}, \mathbf{X}, \bbet^1, \cdots,  \bbet^G) = \sum_{g=1}^{G}  \left\{ \frac{1}{2n}\Vert \mathbf{y} - \mathbf{X} 
		\bbet^g \Vert^{2}_2 + \lambda_{s}P_s\left(\bbet^g\right) +\frac{\lambda_{d}}{2} \sum_{h\neq g}^{G}P_d\left(\bbet^h,\bbet^g\right)\right\},  
		\label{eq:step_reg}
		\end{equation}
		where $G$ is the number of groups, $ P_s(\boldsymbol{\beta })$ is a penalty function that encourages shrinkage and  $P_d\left(\bbet^h,\bbet^g\right)$ is a penalty function that encourages {\it diversity}. The constants $\lambda_s, \lambda_d\geq0 $   determine the amount of shrinkage and diversity, respectively. In this paper we propose the diversity penalty function   
		\begin{equation}
		P_d\left(\bbet^h,\bbet^g\right) = \sum_{j=1}^p |\beta_j^g||\beta_j^h|.
		\label{eq:div_fun}
		\end{equation}
		In our implementation we take 
		$P_{s}$ to be the elastic net penalty 
		\begin{equation} \label{eq:elastic-net}
		P_{s}(\boldsymbol{\beta }^{g})=\left( \frac{(1-\alpha
			)}{2}\Vert \boldsymbol{\beta }^{g}\Vert _{2}^{2}+\alpha \Vert \boldsymbol{%
			\beta}^{g}\Vert_{1}\right) ,
		\end{equation}%
		where $\alpha \in \lbrack 0,1]$.
		
		In Figure \ref{fig:pen_plot} we show level surfaces of the full penalty 
		term for $p=1$, $G=3$, $\alpha=1$,  $\lambda_{s}=1$ and different values of 
		$\lambda_{d}$. Hence the surfaces plotted are the solutions of
		$$
		\vert \beta^{1}_{1}\vert + \vert\beta^{2}_{1}\vert + \vert\beta^{3}_{1} \vert+ 
		\lambda_{d} \left( \vert \beta^{1}_{1} \beta^{2}_{1}\vert + \vert \beta^{1}_{1} 
		\beta^{3}_{1} \vert+ \vert\beta^{3}_{1} \beta^{2}_{1}\vert\right)=1.
		$$
		We see that when $\lambda_{d}$ is small, the surface is similar to the 
		three-dimensional $\ell_1$ ball. For larger values of $\lambda_{d}$ the surface 
		becomes highly non-convex, with peaks aligned with the axes, where there is only 
		one model that is non-null.
		
		\begin{figure}[ht]
			\centering
			\begin{subfigure}[b]{0.3\linewidth}
				\centering\includegraphics[scale=0.35]{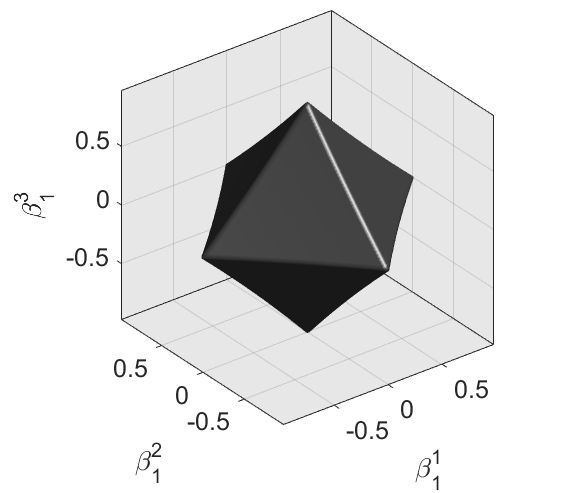}
				\caption{$\lambda_d=0.1$}
			\end{subfigure}%
			\begin{subfigure}[b]{0.3\linewidth}
				\centering\includegraphics[scale=0.35]{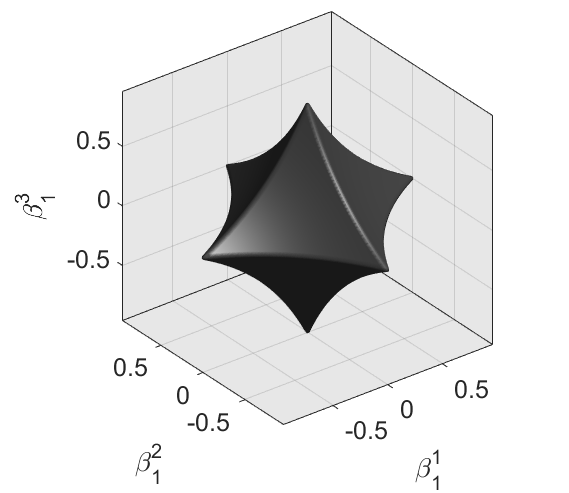}
				\caption{$\lambda_d=1$}
			\end{subfigure}%
			\begin{subfigure}[b]{0.3\linewidth}
				\centering\includegraphics[scale=0.35]{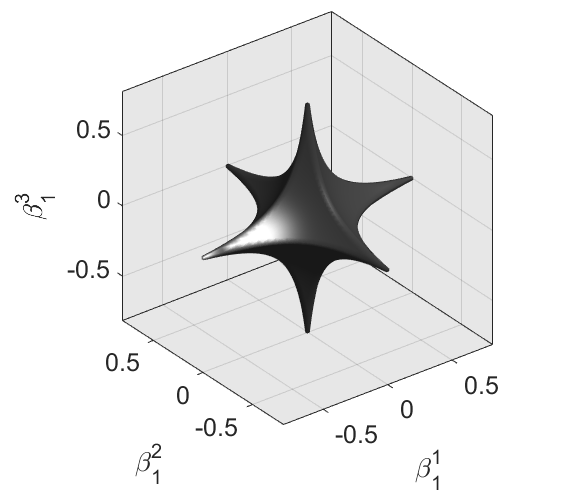}
				\caption{$\lambda_d=10$}
			\end{subfigure}
			\caption{\label{fig:pen_plot}
				Plots of the full penalty term for $\alpha=1$, $\lambda_{s}=1$ and three 
				different values of $\lambda_{d}$.}
		\end{figure}

		
		Let us now draw attention to the following facts. 
		\begin{enumerate}
			\item  $P_d\left(\bbet^h,\bbet^g\right) =0$  if and only if $\beta_j^g\beta_j^h=0 $  for all $j=1,\dots,p$. That is, the variables used in groups $g$ and $h$  are distinct. 
			\smallskip
			
			\item Let $\hbbet =\left( \hbbet^1,\cdots,\hbbet^G \right)$ be a solution of \eqref{eq:step_reg} with $\lambda_{d}=0$ and some $\lambda_{s}$. Then each of the columns of $\hbbet$ is a solution  of the Elastic Net optimization problem with penalty parameter $\lambda_{s}$. This follows immediately from the fact the if $\lambda_{d}=0$, the objective function in \eqref{eq:step_reg} decouples across the different groups, so that each term can be minimized separately.
			
			\smallskip
			
			\item As $\lambda_d \rightarrow \infty$ we have that 
			$P_d\left(\bbet^h,\bbet^g\right) \rightarrow 0$ for all $g \neq h$. That is, active variables in different groups are distinct. 
			\smallskip
		\end{enumerate}
		
		In general, the argument in point 2 above shows that by appropriately choosing the penalty function 
		$P_{s}(\boldsymbol{%
			\beta}^{g})$, our method generalizes penalized regression
		estimators, such as the lasso \citep{Lasso}, the elastic net \citep{EN} and the SCAD \citep{SCAD},  allowing for the selection of possibly overlapping subsets of
		features in such a way that variables that work well together end up in the
		same model, while at the same time encouraging diversity between the models,
		to reduce the correlation between the predictions resulting from each
		of them. This implies that, by appropriately choosing the tuning parameters the proposed
		method automatically and optimally decides: which variables are left out, the distribution of the active variables among the different
		models (with possible overlap) and the amount of shrinkage applied to the active variables
		in each of the models.

		Once we have  obtained the  estimates $\hbbet^{1}, \dots, 
		\hbbet^{G}$ by minimizing (\ref{eq:step_reg}) we must aggregate them to form an overall fit or  prediction by averaging these models: if 
		$\mathbf{x}$ is a training point (new
		observation),  the fitted value (prediction) is given by 
		\begin{equation}
		\widehat{y}(\mathbf{x})= \frac{1}{G}\sum\limits_{g=1}^{G} \mathbf{x}%
		^{\prime} \hbbet^{g}= \mathbf{x}^{\prime} \left( \frac{1}{G}%
		\sum\limits_{g=1}^{G} \hbbet^{g}\right) = \mathbf{x}^{\prime} \hbbet_{*}. 
		\label{eq:avg_model}
		\end{equation}

		Since the output of our method consists of $G$ fitted models to be used together for model fitting and/or prediction, we can view our method as a way of optimal selection of models for ensembling.   
		Therefore, there is a connection  between our method and model ensembling, 
		a powerful  prediction tool. Examples of ensemble
		methods for regression include Random Forests \citep{RF} and Boosting %
		\citep{boosting, GBM}. Both methods can adapt well to the presence of
		non-linearity, but the resulting prediction rules are generally difficult to
		interpret. If the relation between the response and the predictor variables
		is approximately linear, an ensemble of linear models produces highly
		competitive predictions and  yield more interpretable results.
		The groups of variables into which the features are split are not predetermined and  can overlap. The groups are in fact learned from the data through the optimization of the objective function. This is in sharp contrast with the approach in {\em group lasso} \citep{group} where pre-existing domain knowledge determines the assignment of variables to groups.
		
		The problem of minimizing \eqref{eq:step_reg} can be posed as an `artificial' 
		multivariate linear regression problem. Let $\mathbf{Y}\in\mathbb{R}^{n\times 
			G}$ be the matrix with the vector $\mathbf{y}$ repeated $G$ times as columns. 
		$$
		O(\mathbf{y}, \mathbf{X}, \bbet) = \frac{1}{2n}\Vert \mathbf{Y} - \mathbf{X} 
		\bbet\Vert_{F}^{2} + \lambda_{s} \left( \frac{(1-\alpha)}{2}\Vert 
		\bbet\Vert_{F}^{2} + \alpha \Vert \bbet \Vert_1\right)+ \frac{\lambda_{d}}{2} 
		\left( \Vert  \vert\bbet\vert^{\prime} \vert\bbet\vert\Vert_1 - 
		\Vert\bbet\Vert_{F}^{2} \right),
		$$
		where $\Vert \cdot\Vert_{F}$ is the Frobenius norm, $\vert \bbet\vert$ stands 
		for taking the absolute value coordinate-wise and $\Vert \cdot \Vert_{1}$ is the 
		sum of the absolute values of the entries of the matrix. It is seen that the 
		diversity penalty term in a sense penalizes correlations between the different 
		models.

		\subsection{The case of orthogonal predictors}
		
		The objective function in \eqref{eq:step_reg} consists of three terms: a goodness of fit term (the least squares loss), a sparsity penalty term (the $\ell_{1}$ penalty), and the diversity penalty term introduced in this paper. By now the interplay between the least squares loss and the sparsity penalty is well understood. Our goal here is to shed some light on the role played by the diversity penalty in the solutions of \eqref{eq:step_reg}. Unfortunately, the optimization problem that defines the SplitReg estimator does not in general have a closed form solution. However, closed form solutions can be derived for some very simple cases. In the following proposition we derive such closed form solution for the minimizers of the objective function when $G=2$ and the predictors are orthogonal. These simple results give a good indication of the effect that the diversity penalty has on the solutions of \eqref{eq:step_reg}. Namely, Proposition \ref{theo:orth_closed_form} below will show that:
		\begin{enumerate}
			\item If 
			the maximal absolute correlation between the predictors and the response is below
			$\alpha \lambda_{s}$, then all the coefficients of the SplitReg solution are equal to zero. 
			\item On the other hand if 
			the maximal absolute correlation between the predictors and the response exceeds
			$\alpha \lambda_{s}$
			we have essentially two possible distinct regimes depending on the value of the diversity penalty constant $\lambda_{d}$. 
			\begin{enumerate}	
				\item[a)]	
				When  $\lambda_{d} \leq 1 
				+ (1-\alpha)\lambda_{s}$, diversity is not enforced and the features are active in both models.
				\item[b)] When $\lambda_{d} >1 + (1-\alpha)\lambda_{s}$ diversity is fully enforced and the features can be active in at most one of the two models.
			\end{enumerate} 
		\end{enumerate}

		\begin{prop}
			\label{theo:orth_closed_form} Suppose that $G=2$ and  $\mathbf{X}/\sqrt{n}$ is orthogonal. For $j=1,\dots,p$ let 
			$r_{j}=\mathbf{y}^{\prime}\mathbf{x}^{j}/n$. Then
			\begin{enumerate}
				\item If $\vert r_{j} \vert \leq \alpha \lambda_{s}$ the $j$-th coefficients in the two models for all solutions of SplitReg are zero.
				\item If $\vert r_{j} \vert > \alpha \lambda_{s}$
				\begin{enumerate}
					\item If $\lambda_{d} < 1 + (1-\alpha)\lambda_{s}$ all solutions of SplitReg 
					satisfy
					$$
					\hbeta^{1}_{j}=\hbeta^{2}_{j}= \frac{\operatorname{soft}(r_{j}, 
						\alpha\lambda_{s})}{1+(1-\alpha)\lambda_{s}+\lambda_{d}}
					$$ where soft is the soft-thresholding operator,
					defined by $\text{soft}(z, \gamma) = \text{sign}(z)\max(0, \vert z \vert -
					\gamma)$.
					\item If $\lambda_{d} =1 + (1-\alpha)\lambda_{s}$ any pair $(\beta^{1}_{j}, 
					\beta^{2}_{j})$ that satisfies $\beta^{1}_{j} \beta^{2}_{j}\geq 0$ and 
					$$
					\beta^{1}_{j} + \beta^{2}_{j}= \frac{\operatorname{soft}(r_{j}, 
						\alpha\lambda_{s})}{1+(1-\alpha)\lambda_{s}}
					$$
					is a solution to SplitReg.
					\item If $\lambda_{d} > 1 + (1-\alpha)\lambda_{s}$ all solutions of SplitReg 
					satisfy that only one of $\hbeta^{1}_{j}$ and $\hbeta^{2}_{j}$ is zero, and the 
					non-zero one is equal to
					$$
					\frac{\operatorname{soft}(r_{j}, \alpha\lambda_{s})}{1+(1-\alpha)\lambda_{s}}.$$
				\end{enumerate}
			\end{enumerate}
		\end{prop}

		\subsection{The case of two correlated predictors}
		Further insights into how our procedure works can be gained by analyzing the 
		simple case in which there are only two correlated predictors and two models.
		
		\begin{prop}
			\label{theo:corr_close_form}
			Assume $\mathbf{X}\in\mathbb{R}^{n \times 2}$ is normalized so that its columns 
			have squared norm equal to $n$ and $G=2$. Let $\hbbet$ be any solution of SplitReg, $\rho = (\mathbf{x}^{2})^{\prime}\mathbf{x}^{1}/n$ and 
			$r_{j}=\mathbf{y}^{\prime}\mathbf{x}^{j}/n$, $j=1,2$.
			\begin{enumerate}
				\item If the models are disjoint then the active variables in each model have 
				coefficients
				$$
				T_{j}=\frac{\operatorname{soft}(r_{j}, 
					\alpha\lambda_{s})}{1+(1-\alpha)\lambda_{s}},\quad j=1,2,
				$$
				and
				$$
				\lambda_{d}\geq\max\left \lbrace  \frac{\left\vert r_1 - \rho T_{2}\right\vert - 
					\alpha \lambda_{s}}{T_{1}},  \frac{\left\vert r_2 - \rho T_{1}\right\vert - 
					\alpha \lambda_{s}}{T_{2}}\right \rbrace.
				$$
				\item If variable $i$ is inactive in both models, variable $j$ is active in both 
				models and $\lambda_{d}\neq 1 + (1-\alpha)\lambda_{s}$  then the coefficients of 
				variable $j$  are equal to
				$$
				\frac{\operatorname{soft}(r_{j}, \alpha\lambda_{s})}{1+(1-\alpha)\lambda_{s} + 
					\lambda_{d}}.
				$$
				\item Assume $\lambda_{s}=0$ and that both variables are active in both models. 
				If $\sign({\hbeta^{1}_{1}})=\sign({\hbeta^{2}_{1}})$ and $\sign(\hbeta^{1}_{2}) 
				= \sign(\hbeta^{2}_{2})$ then all solutions of SplitReg satisfy
				\begin{equation*}
				\begin{pmatrix}
				1 &\lambda_{d} &0& \rho \\
				\lambda_{d}& 1 &\rho& 0 \\
				0 & \rho & 1 &\lambda_{d} \\
				\rho & 0 & \lambda_{d} & 1
				\end{pmatrix}
				\begin{pmatrix}
				\hbeta^{1}_{1}\\
				\hbeta^{2}_{1}\\
				\hbeta^{2}_{2}\\
				\hbeta^{1}_{2}
				\end{pmatrix}
				=\begin{pmatrix}
				r_{1}\\
				r_{1}\\
				r_{2}\\
				r_{2}
				\end{pmatrix}.
				\end{equation*}
				If $\lambda_{d}<1-\rho$, the solution is unique.
			\end{enumerate}
		\end{prop}
		
		The case in which $\lambda_{s}=0$ is easier to analyze. In this case, the 
		proposition above implies that if the fitted models are disjoint then 
		$
		\lambda_{d} \geq \left\lbrace \left\vert 1 - \rho ({r_1}/{r_2})\right\vert, 
		\left\vert 1 - \rho ({r_2}/{r_1})\right\vert \right\rbrace,
		$
		and the non-null coefficients of SplitReg are equal to the marginal Elastic 
		Net regressions. Note that in the case in which $r_1=r_2$, the size of the 
		diversity penalty required to separate the models decreases as the correlation 
		between the variables increases. 

		\section{Computing SplitReg}
		\label{sec:alg}
		We wish to compute $\hbbet =\left( \hbbet^1,\cdots,\hbbet^G \right)$ that  minimizes
		$$
		O(\mathbf{y}, \mathbf{X}, \bbet^1, \cdots,  \bbet^G) = \sum_{g=1}^{G}  \left\{ \frac{1}{2n} \Vert \mathbf{y} - \mathbf{X} 
		\bbet^g \Vert^{2}_2 + \lambda_{s}P_s\left(\bbet^g\right) + \frac{\lambda_{d}}{2} \sum_{h\neq g}^{G}P_d\left(\bbet^h,\bbet^g\right)\right\},  
		$$
		with
		$
		P_{s}(\boldsymbol{\beta }^{g})=\left( \frac{(1-\alpha
			)}{2}\Vert \boldsymbol{\beta }^{g}\Vert _{2}^{2}+\alpha \Vert \boldsymbol{%
			\beta}^{g}\Vert_{1}\right) ,
		$
		and
		$
		P_d\left(\bbet^h,\bbet^g\right) = \sum_{j=1}^p |\beta_j^g||\beta_j^h|.
		$

		For any $\lambda_{s}>0$, $O(\mathbf{y}, \mathbf{X}, 
		\boldsymbol{\beta}) \to \infty$ as $\Vert \boldsymbol{\beta} \Vert \to \infty
		$ and hence a global minimum of $O$ exists.  But the problem of finding the global minimum, $\hbbet$, is not straightforward because $O
		$ is a non-convex function of $\boldsymbol{\beta}$, due
		to the non-convexity of $P_{d}$. Still we are able to construct an efficient computing algorithm that converges to a coordinate-wise minimum. We first observe that the objective function is convex (strictly if 
		$\alpha<1$) in each coordinate, $\beta^{h}_{j}$,  and in each group of coordinates, $\bbet^{g}$, 
		since the corresponding optimization problems are penalized least 
		squares problems with a weighted elastic net penalty.  In fact, for any fixed  $1\leq g\leq G$ we have
		\begin{align*}
		O(\mathbf{y},\mathbf{X},\boldsymbol{\beta})  & =\frac{1}{2n}\Vert
		\mathbf{y}-\mathbf{X}\boldsymbol{\beta}^{g}\Vert^{2}+\lambda_{s}%
		\frac{(1-\alpha)}{2}\Vert\boldsymbol{\beta}^{g}\Vert_{2}^{2}+\sum
		\limits_{j=1}^{p}|\beta_{j}^{g}|(\lambda_{s}\alpha+\frac{\lambda_{d}}{2}%
		\sum\limits_{h\neq g}|\beta_{j}^{h}|)\\
		& \\
		& =\frac{1}{2n}\Vert\mathbf{y}-\mathbf{X}\boldsymbol{\beta}^{g}\Vert
		^{2}+\lambda_{s}\frac{(1-\alpha)}{2}\Vert\boldsymbol{\beta}^{g}\Vert_{2}%
		^{2}+\sum\limits_{j=1}^{p}|\beta_{j}^{g}|w_{j,g},
		\end{align*}
		where $$w_{j, g}= (\lambda_{s}\alpha  + \lambda_{d}/2 \sum_{h\neq g} \vert 
		\beta^{h}_{j}\vert).$$
		Hence, for each group and for each coordinate separately we have an elastic net type 
		problem, where the lasso penalty has weights which depend on the solution 
		itself. In particular, the coordinates most penalized in model $g$ will be those 
		that have large coefficients in the other models.


		
		\subsection{Computing solutions for fixed $\lambda_s$ and $\lambda_d$ }
		
		Based on the previous discussion, we propose a computing algorithm based on coordinate descent. Coordinate descent  has proven to be efficient
		for  solving regularized least squares problems, see \cite{glmnet} for example. 
		%
		The following result is the main building block of this algorithm.

		\begin{prop}
			\label{prop:CD_updates}
			The coordinate descent update for $\beta^{g}_{j}$ is given by
			\begin{equation*}
			\beta^{new,g}_{j} = 
			\frac{\operatorname{soft}\left(\frac{1}{n}\sum\limits_{i=1}^{n}
				x_{i,j} (y_{i} - y_{i}^{(-j), g}), \alpha\lambda_{s} + \lambda_{d}
				\sum\limits_{h\neq g} \vert \beta^{old,h}_{j}\vert\right)}{1 + (1-\alpha)
				\lambda_{s}},
			\end{equation*}
			where $y_{i}^{(-j), g}$ is the in-sample prediction of $y_i$ using model $g$ 
			and leaving out variable $j$.
		\end{prop}
		Note that the $\ell_1$ shrinkage being applied to the coefficient of variable $j$ in model $g$, 
		$\alpha\lambda_{s} + \lambda_{d}
		\sum_{h\neq g} \vert \beta^{old,h}_{j}\vert$, increases with the sum of the 
		absolute values of the coefficients of variable $j$ in all other models. This 
		shows more clearly that the penalty $ 
		P_{d}$ encourages diversity among the models.
		
		We cycle through the coordinates of $\boldsymbol{\beta}^{1}$,
		then through those of $\boldsymbol{\beta}^{2}$ and so on until we reach 
		$\boldsymbol{%
			\beta}^{G}$, where we check for convergence. Convergence is declared when 
		\begin{equation*}
		\max_{j} \left( \frac{1}{G}\sum\limits_{g=1}^{G}\beta^{new,g}_{j} - \frac{1}{G}%
		\sum\limits_{g=1}^{G}\beta^{old,g}_{j}\right)^{2} < \delta,
		\end{equation*}
		for some small positive $\delta$. Since the
		data is standardized, the convergence criterion in the original units is: 
		\begin{equation*}
		\max_{j} \frac{1}{n}\sum_{i=1}^{n}\left( x_{i,j}\frac{1}{G}%
		\sum\limits_{g=1}^{G}\beta^{new,g}_{j} - x_{i,j}\frac{1}{G}\sum%
		\limits_{g=1}^{G}\beta^{old,g}_{j}\right)^{2} < \delta \frac{1}{n}%
		\sum\limits_{i=1}^{n}\left(y_{i} - \bar{y}\right)^{2}.
		\end{equation*}
		Hence, the algorithm converges when the in-sample average predictions no longer
		change significantly. The following Proposition follows immediately from Theorem 4.1 of \cite{Tseng}
		
		\begin{prop}
			\label{prop:Cord-wise Min}
			The proposed algorithm converges to a coordinate-wise minimizer of \eqref{eq:step_reg}.
		\end{prop}

		%
		
		
		\subsection{Choosing the penalty parameters}
		We choose $\lambda_{s}$ and $\lambda_{d}$ over grids of candidates, looking
		to minimize the cross-validated (CV) mean squared prediction error (MSPE). The grids of
		candidates are built as follows. 
		Let $\lambda_{s}^{max}$ be the smallest $\lambda_{s}$
		that makes all models null. Clearly $\lambda_{s}^{max}$ depends on $\lambda_{d}$.
		It is easy to show that for $\lambda_{d} =
		0$ and $\alpha >0$, $
		\lambda_{s}^{max}= {1}/{(n \alpha)} \max_{j \leq p} \left\vert
		\sum_{i=1}^{n} x_{i,j} y_{i} \right\vert$.
		For $\lambda_{d}>0$ we estimate $
		\lambda_{s}^{max}$ by performing a grid search.
		For each $\lambda_{d}$, the corresponding $\lambda_{s}^{max}$ is the maximum sparsity penalty that will be considered.
		The smallest $\lambda_{d}$ that maximises diversity among the models (makes
		them disjoint) for a given $\lambda_{s}$, say $\lambda_{d}^{max}$, is also
		estimated using a grid search. 
		To build a grid to search for the optimal $%
		\lambda_{s}$ we take 100 log-equispaced points between $\varepsilon
		\lambda_{s}^{max}$ and $\lambda_{s}^{max}$, where $\varepsilon$ is $10^{-4}$
		if $p<n$ and $10^{-2}$ otherwise. The grid used for $\lambda_{d}$ is built
		analogously, but including zero as a candidate.
		
		Even though we could also cross-validate over a grid of possible values of $%
		\alpha$, we find that taking a large value of $\alpha$,
		say $\alpha = 3/4$ or $\alpha=1$, generally works well and hence in what follows 
		we assume
		that $\alpha$ is fixed.
		
		Fix one of $\lambda_{s}$, $\lambda_{d}$. We then minimize the objective
		function $O$ over the grid of candidates corresponding to the other penalty
		term, going from the largest to the smallest values in the grid; for each
		element of the grid of candidates, the solution to the problem using the
		previous element is used as a warm start. Even though the optimal $%
		\hbbet$ is not in general a continuous function of $\lambda_{d}$ and 
		$\lambda_{s}$%
		, see Proposition \ref{theo:orth_closed_form}, we find that using warm starts as
		described above works well in practice.
		
		The main loop of the algorithm works as follows, starting with 
		$\lambda_{d}^{opt} = 0$,
		and until the CV MSPE no longer decreases:
		
		\begin{itemize}
			\item Find the $\lambda_{s}$ in the grid giving minimal CV MSPE, $%
			\lambda_{s}^{opt}$.
			
			\item Take the optimal $\lambda_{s}^{opt}$ from the previous step. Recompute 
			$\lambda_{d}^{max}$ and the corresponding grid. Find the $\lambda_{d}$ in
			the grid giving minimal cross-validated MSPE, $\lambda_{d}^{opt}$. Go to the
			previous step.
		\end{itemize}

		As we mentioned earlier, since we start with $\lambda_{d}^{opt} = 0$, the 
		solution with all
		columns equal to a elastic net estimator is always a candidate to be
		chosen.

		\subsection{The number of models}
		
		We conduct a small simulation study to illustrate the effect increasing the 
		number of models has on the computation time and the performance of SplitReg-Lasso. We generate 100 replications of a linear model with $p=1000$ 
		predictors and $n=100$ observations, corresponding to the second covariance 
		structure described in Section \ref{sec:sim}. For each replication, two 
		data-sets are generated, one on which SplitReg is trained and one used for 
		computing the mean squared prediction error (MSPE). The computation is repeated 
		for various values of the proportion of active variables, called $\zeta$. The signal to noise 
		ratio is 10. We show the MSPEs for different values of the number of models used 
		(rows) and the proportion of active variables in the data generating process 
		(columns). 
		We also computed a measure of the overlap between the models of SplitReg. 
		Let $\hbbet\in\mathbb{R}^{p\times G}$ be the matrix with columns equal to the 
		computed models, where $G$ is the number of models and $p$ the number of 
		features. Let $
		o_j = (1/G)\sum_{g=1}^{G} I\lbrace \hbeta_{j}^{g} \neq 0\rbrace,$
		then we define the overlap as
		$$
		\text{OVP} = \frac{\sum\limits_{j=1}^{p} o_{j} I\lbrace o_j \neq 
			0\rbrace}{\sum\limits_{j=1}^{p} I\lbrace o_j \neq 0\rbrace}
		$$
		if $\sum_{j=1}^{p} I\lbrace o_j \neq 0\rbrace \neq 0$,
		and as 0 otherwise.
		Note that $0 \leq \text{OVP} \leq 1$. If $\text{OVP}=0$ then all models are 
		empty, whereas if $\text{OVP}>0$, then at least one model is non-empty and 
		actually $\text{OVP}\geq 1/G$. If $\text{OVP}=1/G$ then each variable that is 
		active can only appear in one model, and hence the overlap between the models is 
		minimal, since they are disjoint.
		Finally, if $\text{OVP}=1$ then all the variables that are active in at least 
		one model, actually appear in all the models, and hence we have maximum overlap.
		
		Table 1 shows the results.
		In all the considered cases, as the number of models  increases, both the 
		overlap and the MSPE decrease. The gains in prediction accuracy, however,  decrease at $G=10$ and nearly cease passing $G=15$. There seems to be a `diminishing returns' type phenomenon. 
		
		The last column of Table 1 shows the average computation time in seconds. The computation time doesn't 
		vary much for the  different sparsity levels. Hence we report the average 
		over them.  In all the settings studied in this paper, the increase in computational time 
		due to using more models, appears to be approximately linear in the number of 
		models, as evidenced by Table 1. From this table we see that the average computing time $T$ is approximately 3.19+3$G$ seconds.
		
		An objective way to determine a nearly optimal number of models to be used, is to cross-validate 
		over a coarse grid, say, taking $2, 5, 7$ or $10$ models; this is the approach we take
		in Section \ref{sec:real}, where we apply the proposed methodology to a real data-set.
		In our simulations we always use ten models, a possibly sub-optimal choice, but still 
		good enough to give an excellent performance.

		\input{numgroups_results.txt}
		\FloatBarrier

		\section{A consistency result}\label{sec:cons}
		
		Assume the data follows a standard linear regression model
		\begin{equation}
		y_{i}= \mathbf{x}_{i}^{\prime}\bbet_{0} + \varepsilon_{i},\quad 1\leq i \leq n,
		\label{eq:lin_mod}
		\end{equation}
		where the vector of predictors $\mathbf{x}_{i}$ is fixed and the errors $\varepsilon_i$ are
		i.i.d. normal random variables with variance $\sigma^2$.
		The number of predictors $p$ may depend on the sample size and be greater than $n$. As before, we assume that
		$(1/n)\sum\limits_{i=1}^{n} x_{i,j}^{2}=1$ for $j=1\dots p$.
		
		\begin{theorem}\label{theo:cons}
			Let $\alpha\in(0,1]$. Assume that $\lambda_{s}\geq (1/\alpha) \sigma \sqrt{(t^2+2\log(p))/n}$ for some $t>0$. 
			Let $\hbbet$ be any solution of \eqref{eq:step_reg}.
			Then with probability at 
			least $1-2\exp(-t^2/2)$ we have
			$$
			\frac{1}{2n}\left\Vert \left(\frac{1}{G} \sum_{g=1}^{G}\mathbf{X}\hbbet^{g}\right) - \mathbf{X}\bbet_0 \right\Vert_{2}^{2} \leq 2\alpha \lambda_{s} \Vert \bbet_0\Vert_{1} + \lambda_{s}\frac{(1-\alpha)}{2}\Vert \bbet_{0}\Vert_{2}^{2}+ \frac{\lambda_{d} (G-1)}{2} \Vert \bbet_0 \Vert_{2}^2.
			$$
		\end{theorem}
		
		It follows that if we take $\lambda_{s}$ to be of order $\sqrt{\log(p)/n}$ and $\lambda_{d}$ to be of order
		$\log (p)/n$ then if we assume $\Vert \bbet_0 \Vert_{1}$ and $\Vert \bbet_0 \Vert_{2}^{2}$ are of order smaller than $\sqrt{n/\log(p)}$ and
		$\log(p)/n \to 0$, the average prediction, 
		$(1/G)\sum_{g=1}^{G} \mathbf{X}\hbbet^{g}$, is consistent.
		A similar result can be obtained if one assumes only that the errors have a sub-gaussian distribution.
		Sharper bounds may be obtained if one assumes more restrictive conditions on the set of predictor variables, for
		example the so-called \emph{compatibility condition};
		see Section 6.2.2 of \cite{buhlmann-book} for details. An overview of consistency results for regularized estimators is available in, for example, \cite{buhlmann-book}.
		In the more classical case in which the smallest eigenvalue of $\mathbf{X}^{\prime}\mathbf{X}/n$ is bounded below by a fixed constant,
		for example when $p$ is taken to be fixed and $\mathbf{X}^{\prime}\mathbf{X}/n$ converges to a positive definite matrix, we may deduce that
		$\Vert (1/G) \sum_{g=1}^{G}\hbbet^{g} - \bbet_0 \Vert_{2}^{2} \to 0$ in probability as $n$ goes to infinity.	
		
		Theorem \ref{theo:cons} applies only to global solutions of \eqref{eq:step_reg}, which we have no guarantees coincide with the output provided by our algorithm. However, the asymptotic analysis of the output of the algorithm seems unfeasible.

		\section{Simulations}\label{sec:sim}
		
		\subsection{Methods}
		
		An \texttt{R} package that implements the procedures proposed in this paper, called \texttt{SplitReg} is available from \texttt{CRAN}. We ran a simulation study, comparing the prediction accuracy of the following 
		eleven competitors. All computations were carried out in \texttt{R}. 
		\begin{enumerate}
			\item {\bf  Lasso}, computed using the \texttt{glmnet} package.
			
			\item {\bf Elastic net} with $\alpha = 3/4$, computed using the 
			\texttt{glmnet} package.
			
			\item Split regularized regression  using $G=10$ and lasso sparsity penalty, called {\bf SplitReg-Lasso}.
			
			\item Split regularized regression  using $G=10$ and elastic net  sparsity penalty, called {\bf SplitReg-EN}.

			\item Sure independence screening (SIS) procedure, \cite{SIS}, followed
			by fitting a SCAD penalized least squares estimator, computed using the 
			\texttt{SIS} package, called {\bf SIS-SCAD}.
			
			\item MC+ penalized least squares estimator, \cite{MCP}, computed using the 
			\texttt{sparsenet} package, called {\bf SparseNet}.
			
			\item Relaxed Lasso, \cite{relaxed}, computed using the \texttt{relaxnet} 
			package, called {\bf Relaxed}.
			
			\item Forward stepwise algorithm, computed using the \texttt{lars} package, 
			called {\bf Stepwise}. 
			
			\item Cluster representative lasso, proposed in \cite{buhlmann}, computed 
			using code kindly provided by the authors, called {\bf CRL}.
			
			\item Random forest of \cite{RF}, computed using the \texttt{randomForest} package, called
			\textbf{RF}.
			
			\item Random GLM method of \cite{rglm}, computed using the \texttt{randomGLM} package,
			called \textbf{RGLM}.
		\end{enumerate}
		
		
		All tuning parameters were chosen via cross-validation; the RF and RGLM methods
		were ran using their default settings. The CRL of 
		\cite{buhlmann} was not included in scenarios with $p=1000$ due to its long 
		computation time when compared with the rest of the methods. For the same 
		reason, in the scenarios with $p=150$, we only did 100 replications for CRL, 
		instead of the 300 done for all the other procedures. 
		
		The popular group lasso \cite{group, sparse-group} is not included in the 
		simulation, because we don't assume that there is a priori knowledge of the 
		existence of pre-defined groups among the features.
		The interesting recent proposals of
		\cite{buhlmann}, \cite{PACS} and \cite{witten}, 
		assume that there exist unknown clusters of correlated variables, and shrink the 
		coefficients of variables in the same cluster towards each other. Because  CRL 
		has a relatively more efficient numerical implementation (compared with the 
		other two) we included it in our simulation to represent the cluster-based 
		approaches.
		Finally, note that all the competitors above, except perhaps for the forward 
		stepwise algorithm, the RF and the RGLM, could in principle be used as building blocks in our 
		procedure. One of the main objectives of this simulation study is to show that 
		the proposed method for splitting the features improves upon the prediction accuracy 
		of the base estimators being pooled together, in this case, the Lasso and the Elastic Net.

		\subsection{Models}
		
		For each Monte Carlo replication, we generate data from a linear
		model: 
		\begin{equation*}
		y_{i} = \mathbf{x}_{i}^{\prime} \boldsymbol{\beta}_{0} + \sigma \epsilon_{i},
		\quad 1\leq i \leq n,
		\end{equation*}
		where the $\mathbf{x}_{i}\in\mathbb{R}^{p}$ are multivariate normal with
		zero mean and correlation matrix $\boldsymbol{\Sigma}$ and the $\epsilon_{i}$
		are standard normal. We consider two combinations of $p$ and $n$, namely $(p,n)=(1000, 100)$ and $(p,n)=(150, 75)$.
		For each $p$, we take the number of active variables to be $p_0=[p \zeta]$
		with $\zeta = 0.05, 0.1, 0.2$ and $0.4$. 
		
		For all scenarios, the nonzero elements of the $p$-dimensional vectors $\boldsymbol{\beta}_0$ are randomly generated as described in \cite{SIS}. We set $a = 5 \log n/\sqrt{n}$, and set nonzero coefficients to be $ (-1)^u (a + |z|) $ for each model, where $u$ was drawn from a Bernoulli distribution with parameter 0.2 and $ z $ is drawn from the standard Gaussian distribution. For $(p,n)=(1000, 100)$, the $ l_2 $-norms $ \Vert \boldsymbol{\beta}_{0} \Vert $ range from 18.50 to 54.80 for all sparsity levels $1 - \zeta$ considered. For $(p,n)=(150, 75)$, the $ l_2 $-norms $ \Vert \boldsymbol{\beta}_{0} \Vert $ range from 7.94 to 21.13.
		
		Given $p$, $n$ and a sparsity level $1 - \zeta$, the following scenarios for $\boldsymbol{\beta}_0$ and $\boldsymbol{\Sigma}$ are considered
		
		\begin{itemize}
			\item[] \textbf{Scenario 1} $\Sigma_{i,j}=\rho$ for all $i\neq j$, the first $[p_0 \zeta]$ coordinates of $\boldsymbol{\beta}_0$ are nonzero and the rest are equal to zero.
			
			\item[] \textbf{Scenario 2}
			\begin{equation*}
			\Sigma_{i,j}=  
			\begin{cases}
			1 & \text{if } i=j \\ 
			\rho & \text{if } 1\leq i,j \leq \lfloor p_0/2 \rfloor + \lceil (p-p_0)/2\rceil, 
			i\neq j \\ 
			\rho & \text{if } \lfloor p_0/2 \rfloor + \lceil (p-p_0)/2\rceil + 1 \leq i,j 
			\leq p, i\neq j \\
			0 & \text{otherwise}
			\end{cases}
			\end{equation*}
			$\beta_{j}\neq 0$ for $j\leq \lfloor p_0/2 \rfloor$, $\beta_{j}\neq 0$ for $\lfloor 
			p_0/2 \rfloor + \lceil (p-p_0)/2\rceil
			+ 1 \leq j \leq \lfloor p_0/2 \rfloor + \lceil (p-p_0)/2\rceil + p_0 - \lfloor 
			p_0/2 \rfloor$ and the rest of the
			coordinates equal to zero.
			
			\item[] \textbf{Scenario 3} \begin{equation*}
			\Sigma_{i,j}=  
			\begin{cases}
			1 & \text{if } i=j \\ 
			\rho & \text{if } 1\leq i,j \leq  p_0, i\neq j \\ 
			0 & \text{otherwise},
			\end{cases}
			\end{equation*}
			the first $[p_0 \zeta]$ coordinates of $\boldsymbol{\beta}_0$ are nonzero and the rest are zero.
		\end{itemize}
		We consider different values of $\rho$: $0.2$, $0.5$, $0.8$.
		Then $\sigma$ is chosen to give a desired signal to noise ratio (SNR), defined 
		as $\text{SNR} = {\bbet_{0}^{\prime} \boldsymbol{\Sigma} \bbet_{0}}/{\sigma^2}.$
		We consider SNRs of 3, 5 and 10.
		In Scenario 1, all the predictors are correlated among each other. In Scenario 2, we have two groups of active variables, similar to the simulation scenario considered in \cite{witten}. Variables within each group are 
		correlated with each other, but the groups are independent. In Scenario 3, the active variables are only correlated with each other.
		We report results for all scenarios across all considered correlations, SNRs and the two combinations of $p$ and $n$.
		
		\subsection{Performance measures}
		For each replication, two independent copies of the data are generated, one
		to fit the procedures, the other one to compute the MSPE, divided by the variance of the noise, $\sigma^2$. Hence, the best 
		possible result is $1$. 
		In each table reporting the MSPEs, we also compute the standard error for each of 
		the methods, and report the maximum among them in the caption. We also compute and report the precision (PR) and recall (RC) of each method, defined as
		\begin{align*}
		&\text{PR} = \frac{\# \lbrace j:\beta_{0,j} \neq 0 \wedge \beta_j \neq 0 
			\rbrace}{\# \lbrace j: \beta_j \neq 0 \rbrace}, \quad \text{RC} = \frac{\# \lbrace j:\beta_{0,j} \neq 0 \wedge \beta_j \neq 0 
			\rbrace}{\# \lbrace j:\beta_{0,j} \neq 0 \rbrace}.
		\end{align*}
		Note that large values of PR and RC are desirable.  For SplitReg, the vector of coefficients used to compute the precision and 
		recall is the average of the models, see \eqref{eq:avg_model}. For the SIS-SCAD 
		method, the precision and recall are computed using the variables selected by 
		the SIS step. For the RGLM method, the precision and recall are computed using the
		union of the variables selected in each of the bags. Since RF does not fit a linear model,
		we do not compute its precision and recall.
		
		\subsection{Results}
		In Table \ref{tab:rank} we report the average rank for all the compared methods  with the exception of CRL because this method was only evaluated in the case $ (p,n)=(150,100)$.   Rank 1 corresponds to the top performer  method and rank 10 corresponds to  the worst performing method. The average ranks are  evaluated across  all the considered scenarios, correlations, SNRs and sparsity levels. We report the average ranks  for each of  the two considered combinations of $p$ and $n$,  and the overall average ranks. The average ranks of the  best two performing  methods are displayed in bold face.
		
		For $ (p,n)=(1000,100) $, SplitReg-EN and SplitReg-Lasso have the best average ranks for the MSPE, with RGLM being the runner up. We also note that the average recall ranks of SplitReg-Lasso and SplitReg-EN are better than those of the corresponding base estimators Lasso and elastic net, respectively, and also better than  the other competitors, except for  RGLM. The price paid by RGLM for this 
		improvement is a severe decrease in precision, as RGLM has the worst average rank regarding  precision.
		
		For $ (p,n)=(150,75) $, SplitReg-EN and SplitReg-Lasso again have the best average MSPE ranks. However, RGLM is no longer the closest  competitor as it now displays  the worst average MSPE rank. RGLM also has the worst average rank for the precision, indicating that too many noise variables were selected by the algorithm, deteriorating its performance. The average rank for the recall of SplitReg-Lasso and SplitReg-EN is again better than that of the base estimators Lasso and elastic net. 
		
		Regarding  the overall average ranks, SplitReg-EN and SplitReg-Lasso have the best MSPE performance,   RGLM is the clear winner regarding  recall rank. But this improvement of  RGLM in terms of recall rank is at the cost of an important loss in precision, particularly in the case of 
		$(p,n)=(150,75)$. The relaxed Lasso and forward stepwise regression typically have the best overall precision ranks, but the methods are not competitive regarding  MSPE and recall.
		
		In the scenarios we consider, Elastic Net and Lasso behave very similarly, as do SplitReg-EN and SplitReg-Lasso. This is somewhat surprising specially in cases where the correlation among the variables is high, since this is where the elastic net penalty  would be expected to perform better than $ L_1 $-regularization.  
		\begin{table}[ht]
			\caption{Average rank of the methods over all scenarios, correlations, SNRs and sparsities for $(p,n)=(1000,100)$ and $(p,n)=(150,75)$. The last column contains the overall rank over both combinations of $(p,n)$.} 
			\centering
			\begin{tabular}{lrrrrrrrrr}
				\hline & \multicolumn{3}{c}{$\mathbf{\boldsymbol{p}=1000}$} & \multicolumn{3}{c}{$\mathbf{\boldsymbol{p}=150}$} & \multicolumn{3}{c}{\textbf{Overall Rank}} \\ \cmidrule(lr){2-4} \cmidrule(lr){5-7} \cmidrule(lr){8-10} \textbf{Method} & \textbf{MSPE} & \textbf{RC} & \textbf{PR} & \textbf{MSPE} & \textbf{RC} & \textbf{PR} & \textbf{MSPE} & \textbf{RC} & \textbf{PR} \\ \cmidrule(lr){1-1} \cmidrule(lr){2-4} \cmidrule(lr){5-7} \cmidrule(lr){8-10}Lasso & 5.11 & 5.80 & 4.53 & 2.86 & 3.19 & 3.35 & 3.99 & 4.50 & 3.94 \\ 
				Elastic Net & 3.93 & 4.56 & 4.30 & 2.68 & 2.79 & 3.70 & 3.31 & 3.67 & 4.00 \\ 
				SplitReg-Lasso & \textbf{2.09} & 2.82 & 6.73 & \textbf{1.67} & 1.93 & 4.17 & \textbf{1.88} & 2.38 & 5.45 \\ 
				SplitReg-EN & \textbf{1.59} & \textbf{2.18} & 6.61 & \textbf{1.93} & \textbf{1.83} & 4.43 & \textbf{1.76} & \textbf{2.00} & 5.52 \\ 
				SparseNet & 6.03 & 4.70 & \textbf{3.04} & 3.14 & 3.81 & 2.53 & 4.58 & 4.25 & 2.79 \\ 
				Relaxed & 6.60 & 7.09 & \textbf{2.39} & 3.28 & 4.92 & \textbf{1.49} & 4.94 & 6.00 & \textbf{1.94} \\ 
				Stepwise & 9.51 & 9.00 & 3.37 & 5.00 & 5.89 & \textbf{1.27} & 7.25 & 7.45 & \textbf{2.32} \\ 
				RF & 7.49 & $ - $ & $ - $ & 4.76 & $ - $ & $ - $ & 6.12 & $ - $ & $ - $ \\ 
				RGLM & 3.45 & \textbf{1.00} & 8.35 & 6.47 & \textbf{1.00} & 6.00 & 4.96 & \textbf{1.00} & 7.17 \\ 
				SIS-SCAD & 9.19 & 7.84 & 5.69 & 4.89 & 4.75 & 3.06 & 7.04 & 6.29 & 4.38 \\ 
				\hline
			\end{tabular}
			\label{tab:rank}
		\end{table}

		\FloatBarrier
		\afterpage{\clearpage}
		\section{Glass data-sets}\label{sec:real}
		
		We analyze the performance of the competitors considered in the previous section 
		when predicting on real data-sets from a chemometric problem. To evaluate the prediction 
		accuracy of the competitors we randomly split the data into a training set that 
		has $50\%$ of the observations and a testing set that has the remaining $50\%$. 
		This is repeated 100 times and the resulting MSPEs are averaged. The 
		results are reported relative to the best average performance among all 
		estimators. Hence, the estimator with the best average performance will have a 
		score of 1. We also report the average rank among the data-sets for each method.
		
		The glass data-sets \cite{glass} were obtained from an electron probe X-ray
		microanalysis (EPXMA) of archaeological glass samples. A spectrum on 1920 
		frequencies was measured on a total of 180 glass samples. The goal is to predict
		the concentrations of the following chemical compounds using the spectrum:
		Na2O, MgO, Al2O3, SiO2, P2O5, SO3, Cl, K2O, CaO, MnO, Fe2O3, BaO and PbO.
		After 
		removing predictors with little variation, we are left with $p=486$ frequencies 
		and $n=180$ observations. The CRL estimator was not included in the comparison, due to its long computation time.
		The number of models used to form the ensembles is chosen by cross-validation
		among the values 2, 5, 7, 10. The elastic net and SplitReg-EN
		were computed with $\alpha=0.1$, closer to a Ridge than a Lasso estimator, since
		we a priori expected a relatively low level of sparsity.

		Table \ref{tab:glass} shows the results. Highlighted in black is the best performing method for 
		each compound. It is seen that SplitReg-Lasso has the best overall behavior, 
		having the highest average rank (1.92) over the thirteen compounds. Excluding SplitReg-EN which performs similarly to SplitReg-Lasso, the RGLM is the strongest competitor, with
		a rank of 3.54.
		\begin{table}[ht]
			\caption{Average MSPEs for each compound over 100 random splits into training and testing sets. Last column shows the average rank over all compounds.} 
			\centering
			\resizebox{\textwidth}{!}{
				\begin{tabular}{l|rrrrrrrrrrrrr|r}
					\hline
					\textbf{Method} & \textbf{Na2O} & \textbf{MgO} & \textbf{Al2O3} & \textbf{SiO2} & \textbf{P2O5} & \textbf{SO3} & \textbf{Cl} & \textbf{K2O} & \textbf{CaO} & \textbf{MnO} & \textbf{Fe2O3} & \textbf{BaO} & \textbf{PbO} & \textbf{Rank} \\ 
					\hline
					Lasso & 1.70 & 1.16 & 1.38 & 2.78 & 1.19 & 1.14 & 1.39 & 1.41 & 1.48 & 1.09 & 1.38 & 1.04 & 1.01 & 4.23 \\ 
					Elastic Net & 1.62 & 1.18 & 1.74 & 2.73 & 1.31 & 1.17 & 1.13 & 1.44 & 1.47 & 1.05 & 1.23 & 1.01 & 1.08 & 4.62 \\ 
					SplitReg-Lasso & 1.31 & \textbf{1.00} & \textbf{1.00} & 1.67 & \textbf{1.00} & \textbf{1.00} & \textbf{1.00} & 1.36 & 1.40 & 1.05 & 1.26 & 1.03 & \textbf{1.00} & \textbf{1.92} \\ 
					SplitReg-EN & 1.58 & 1.18 & 1.66 & 2.56 & 1.25 & 1.12 & 1.10 & 1.43 & 1.47 & 1.04 & 1.20 & \textbf{1.00} & 1.08 & 3.38 \\ 
					SparseNet & 1.63 & 1.31 & 1.76 & 2.84 & 1.28 & 1.18 & 1.54 & 1.53 & 1.57 & 1.15 & 1.42 & 1.06 & 1.06 & 6.31 \\ 
					Relaxed & 1.80 & 1.23 & 1.40 & 2.81 & 1.16 & 1.19 & 1.38 & 1.40 & 1.49 & 1.13 & 1.46 & 1.10 & 1.28 & 5.62 \\ 
					Stepwise & 2.40 & 2.13 & 3.72 & 4.10 & 2.65 & 1.52 & 4.75 & 1.89 & 1.78 & 1.18 & 1.51 & 2.32 & 1.73 & 9.00 \\ 
					RF & \textbf{1.00} & 1.95 & 7.04 & 3.02 & 16.48 & 1.16 & 12.06 & 6.93 & 9.18 & 1.16 & \textbf{1.00} & 2.69 & 6.34 & 7.77 \\ 
					RGLM & 1.80 & 1.02 & 1.20 & \textbf{1.00} & \textbf{1.00} & 1.26 & 1.62 & \textbf{1.00} & \textbf{1.00} & \textbf{1.00} & 1.19 & 1.72 & 1.04 & 3.54 \\ 
					SIS-SCAD & 2.01 & 4.19 & 1.70 & 2.88 & 1.85 & 1.29 & 1.87 & 2.08 & 2.08 & 1.19 & 1.67 & 1.68 & 1.95 & 8.62 \\ 
					\hline
				\end{tabular}
			}
			\label{tab:glass}
		\end{table}

		\FloatBarrier
		
		\section{Discussion}
		
		\label{sec:disc}
		
		We have proposed a novel method for splitting variables in linear regression models. 
		The seminal work by \cite{RF} points out that the performance of a given ensemble of models  depends on the strength of the models and their diversity. We believe that the good performance of SplitReg derives from the fact that the loss function underlying  this procedure encourages a good balance between the  models strength and their diversity.  
		In the limiting case when the penalty parameters for the diversity penalty, $\lambda_{d}$, is equal to zero, all the columns of the matrix
		$\widehat{\bbet}$ are identical and equal to the elastic net solution. In the
		opposite $\lambda_{d}=\infty$ limiting case there is at most one non-zero
		element on each row of the matrix $\widehat{\bbet}.$ In practice, the value of
		$\lambda_{d}$ is chosen by cross-validation and the result is a compromise
		between these two extremes. Another factor that may influence the performance of a given ensemble is the weight assigned to each model.  This is not optimized by SplitReg where all the models are given equal weight. On the other hand, the paper by \cite{ando} do not optimize the generation of the models,  putting the accent on the assignment of the weights.  Therefore, their work and ours are complementary.
		
		Examples using real and synthetic data-sets show that the approach 
		systematically improves the prediction accuracy of the base estimators being 
		pooled together. In the synthetic data-sets, the improvements tend to increase with 
		the signal to noise ratio and the number of active variables. 
		For low values of the signal to noise ratio, or when the number of active variables is small, the improvements are negligible. We believe that
		the results reported in this paper show that the proposed method is a valuable
		addition to the practitioners toolbox.

		The approach taken in this paper can be
		extended in several ways. 
		Other sparsity penalties such as the SCAD can be handled similarly.  In fact,  
		the algorithm proposed here will work with any regularized model approach 
		provided the coordinate descent updates can be expressed in closed form. Our 
		method can be extended to GLMs by splitting variables for GLM estimators instead of linear regression
		estimators. For example, groups of logistic regression models can be formed 
		by replacing the 
		quadratic loss in \eqref{eq:step_reg} with the deviance.
		The method can be robustified to deal with outliers by using, for example, a 
		bounded loss function to measure the goodness of fit of each model in 
		\eqref{eq:step_reg}, instead of the classical least squares loss; in this
		case regularized robust regression estimators (see \cite{SY17} for example)
		would be formed. Lower computational times may be achieved by using early 
		stopping strategies when computing solution paths over one of  the penalties and 
		also by using an active set strategy when cycling over the groups, see 
		\cite{glmnet}.
		
		
		\bigskip
		\begin{center}
			{\large\bf Acknowledgments}
		\end{center}
		Part of this work was conducted while Ezequiel Smucler was a postdoctoral research fellow at the departments of Statistics and of Computer Science at The University of British Columbia.
		
		\bigskip
		\begin{center}
			{\large\bf SUPPLEMENTARY MATERIAL}
		\end{center}
		The supplemental material available contains the proofs of all results stated
		in the paper and the full results of our simulation study. An \texttt{R} package that implements the procedures proposed in this paper, called \texttt{SplitReg} is available from \texttt{CRAN}.
		
		\bibliography{SplitRegRef}
		
		\section*{Appendix A: Proofs of the results in the main paper}
		\begin{proof}[Proof of Proposition 1]
			Consider any solution of SplitReg, $\hbbet$, and fix $1\leq j_0 \leq p$. 
			Let $s_{R} = 1 + (1-\alpha)\lambda_{s}$.
			The orthogonality of $\mathbf{X}/\sqrt{n}$ implies that
			\begin{align*}
			\hbeta^{1}_{j_0} = \frac{\text{soft}(r_{j_0}, \lambda_{d} \vert \hbeta^{2}_{j_0}\vert 
				+ \alpha\lambda_{s})}{s_{R}}, \quad \hbeta^{2}_{j_0} = \frac{\text{soft}(r_{j_0}, \lambda_{d} \vert 
				\hbeta^{1}_{j_0}\vert + \alpha\lambda_{s})}{s_{R}}.
			\end{align*}
			Hence if $\vert r_{j_0} \vert \leq \alpha \lambda_{s}$, 
			$\hbeta^{1}_{j_0}=\hbeta^{2}_{j_0}=0$. This proves 1.
			
			Assume now  and until the end of this proof that $\vert r_{j_0} \vert > \alpha 
			\lambda_{s}$. Moreover, assume for now that $r_{j_0}>0$. The case $r_{j_0}<0$ is 
			dealt with similarly. 
			Since $r_{j_0} >  \alpha \lambda_{s}$, it can't happen that $\hbeta^{1}_{j_0}= 
			\hbeta^{2}_{j_0}=0$.
			Assume that only one of $\hbeta^{1}_{j_0}=0$ or $\hbeta^{2}_{j_0}=0$ holds. Without 
			loss of generality, we can assume $\hbeta^{2}_{j_0}=0$. Hence
			\begin{align*}
			\hbeta^{1}_{j_0} = \frac{r_{j_0}-\alpha \lambda_{s}}{s_{R}}
			, \quad \hbeta^{2}_{j_0} = \frac{\left(r_{j_0} - \lambda_{d} \hbeta^{1}_{j_0} - \alpha 
				\lambda_{s} \right)^{+}}{s_{R}}= 0,
			\end{align*}
			so that it must be that 
			$r_{j_0}\leq \lambda_{d} \hbeta^{1}_{j_0}+\alpha\lambda_{s}$ and hence that
			$r_{j_0} - \alpha\lambda_{s} \leq 
			(\lambda_{d}/s_{R})(r_{j_0}-\alpha\lambda_{s}),$
			which implies $\lambda_{d} \geq s_{R}$. We have thus shown that if $\lambda_{d}  
			< s_{R}$, $\hbeta^{1}_{j_0}$ and $\hbeta^{2}_{j_0}$ are both non-zero. Now assume 
			both $\hbeta^{1}_{j_0}$ and $\hbeta^{2}_{j_0}$ are non-zero and $\lambda_{d}\neq 
			s_{R}$. Then
			\begin{align*}
			&\hbeta^{1}_{j_0} = \frac{r_{j_0}- \lambda_{d}  \hbeta^{2}_{j_0}- 
				\alpha\lambda_{s}}{s_{R}} , \hbeta^{2}_{j_0} = \frac{r_{j_0}- \lambda_{d}  
				\hbeta^{1}_{j}- \alpha\lambda_{s}}{s_{R}}\Rightarrow 
			\\ &s_{R}^{2} \hbeta^{1}_{j_0} = s_{R}(r_{j_0} - \alpha\lambda_{s}) - 
			\lambda_{d} r_{j_0}+\lambda_{d}^{2}\hbeta^{1}_{j_0}+\alpha\lambda_{s}\lambda_{d}.
			\end{align*}
			Solving for $\hbeta^{1}_{j_0}$ gives 2(a) for $\hbeta^{1}_{j_0}$, the same 
			argument proves it for $\hbeta^{2}_{j_0}$.
			
			Using the orthogonality of $\mathbf{X}/\sqrt{n}$ it is easy to show that for any 
			$\bbet$
			\begin{align*}
			O(\mathbf{y}, \mathbf{X}, \bbet) &= \frac{1}{2n} \sum\limits_{i=1}^{n}y_{i}^{2}+ 
			\frac{1}{2} \sum\limits_{j=1}^{p} (\beta^{1}_{j})^{2} - \sum\limits_{j=1}^{p} 
			\beta^{1}_{j}r_{j} + \frac{(1-\alpha)\lambda_{s}}{2} \sum\limits_{j=1}^{p} 
			(\beta^{1}_{j})^{2} + \alpha \lambda_{s} \sum\limits_{j=1}^{p} 
			\vert\beta^{1}_{j}\vert
			\\ & + \frac{1}{2n}\sum\limits_{i=1}^{n}y_{i}^{2}+ \frac{1}{2} 
			\sum\limits_{j=1}^{p} (\beta^{2}_{j})^{2} - \sum\limits_{j=1}^{p} 
			\beta^{2}_{j}r_{j} + \frac{(1-\alpha)\lambda_{s}}{2} \sum\limits_{j=1}^{p} 
			(\beta^{2}_{j})^{2} + \alpha \lambda_{s} \sum\limits_{j=1}^{p} 
			\vert\beta^{2}_{j}\vert
			\\& +\lambda_{d} \sum\limits_{j=1}^{p}\vert \beta^{1}_{j}\vert \vert 
			\beta^{2}_{j}\vert .
			\end{align*}
			In particular, if as before $\hbbet$ is any solution to SplitReg, we 
			have that $\vert \hbeta^{1}_{j} \hbeta^{2}_{j} \vert = \hbeta^{1}_{j} 
			\hbeta^{2}_{j}$ for all $j$ and hence $\hbbet$ minimizes
			\begin{align*}
			\sum\limits_{j=1}^{p} \left( s_{R}\left( (\beta^{1}_{j})^{2} + 
			(\beta^{2}_{j})^{2}\right)+2\alpha\lambda_{s}(\vert {\beta}^{1}_{j} \vert + 
			\vert  {\beta}^{2}_{j} \vert)  + 2\lambda_{d} \beta^{1}_{j} \beta^{2}_{j} - 
			2\left( \beta^{1}_{j} + \beta^{2}_{j}\right)r_{j}\right).
			\end{align*}
			If $\lambda_{d}=s_{R}$, we get
			\begin{align*}
			\sum\limits_{j=1}^{p} \left( s_{R}\left(\beta^{1}_{j} + 
			\beta^{2}_{j}\right)^{2}+2\alpha\lambda_{s}(\vert {\beta}^{1}_{j} \vert + \vert 
			{\beta}^{2}_{j} \vert)  - 2\left( \beta^{1}_{j} + 
			\beta^{2}_{j}\right)r_{j}\right).
			\end{align*}
			Assume $\hbeta^{1}_{j_0}$ is the non-null coordinate of SplitReg. If 
			$\hbeta^{1}_{j_0} > 0$ we have
			\begin{align*}
			&s_{R}\left(\hbeta^{1}_{j_0} + \hbeta^{2}_{j_0}\right)^{2}+2\alpha\lambda_{s}(\vert 
			{\hbeta}^{1}_{j_0} \vert + \vert {\hbeta}^{2}_{j_0} \vert)  - 2\left( \hbeta^{1}_{j_0} 
			+ \hbeta^{2}_{j_0}\right)r_{j_0} 
			\\ &= s_{R}\left(\hbeta^{1}_{j_0} + 
			\hbeta^{2}_{j_0}\right)^{2}+2\alpha\lambda_{s}({\hbeta}^{1}_{j_0}  + 
			{\hbeta}^{2}_{j_0} )  - 2\left( \hbeta^{1}_{j_0} + \hbeta^{2}_{j_0}\right)r_{j_0}
			\end{align*}
			which is minimized whenever $\hbeta^{1}_{j_0} + \hbeta^{2}_{j_0} = (r_{j_0} - \alpha 
			\lambda_{s})/s_{R}$; if $\hbeta^{1}_{j_0} < 0$ the corresponding expression is 
			minimized whenever $\hbeta^{1}_{j_0} + \hbeta^{2}_{j_0} = (r_{j_0} + \alpha 
			\lambda_{s})/s_{R}$. This proves 2(b).
			
			Now take $\lambda_{d}  > s_{R}$. Assume that there exists a solution of SplitReg $\hbbet$ such that for $j\in \mathcal{J}=\lbrace j_1,\dots, j_k 
			\rbrace$, $k \geq 1$, both $\hbeta^{1}_{j}$ and $\hbeta^{2}_{j}$ are non-zero, 
			whereas for $j \in \mathcal{J}^{c}$, $j=1,\dots,p$ at most one of 
			$\hbeta^{1}_{j}$ and $\hbeta^{2}_{j}$ is non-zero. Take $\bbet_{\ast}$  equal to 
			$\hbbet$ except that for $j\in\mathcal{J}$, coordinate $(j, 1)$ is equal to 
			$\text{soft}(r_{j}, \alpha \lambda_{s})/ s_{R}$ and coordinate $(j, 2)$ is equal 
			to zero.
			Note that
			\begin{align*}
			O(\mathbf{y}, \mathbf{X}, \hbbet) - O(\mathbf{y}, \mathbf{X}, \bbet_{*}) &= 
			\frac{1}{2} \sum\limits_{j=1}^{p} \left( (\hbeta^{1}_{j})^{2} - 
			(\beta^{1}_{*,j})^{2} \right) + \sum\limits_{j=1}^{p} \left( 
			\beta^{1}_{*,j}r_{j} - \hbeta^{1}_{j}r_{j}\right)
			\\ &+ \frac{(1-\alpha)\lambda_{s}}{2}  \sum\limits_{j=1}^{p} \left( 
			(\hbeta^{1}_{j})^{2} - (\beta^{1}_{*,j})^{2} \right) 
			\\ &+ \alpha \lambda_{s} \sum\limits_{j=1}^{p} \left(\vert\hbeta^{1}_{j}\vert - 
			\vert \beta^{1}_{*,j}\vert \right)
			\\ &+ \frac{1}{2} \sum\limits_{j=1}^{p} \left( (\hbeta^{2}_{j})^{2} - 
			(\beta^{2}_{*,j})^{2} \right) + \sum\limits_{j=1}^{p} \left( 
			\beta^{2}_{*,j}r_{j} - \hbeta^{2}_{j}r_{j}\right)
			\\ &+ \frac{(1-\alpha)\lambda_{s}}{2}  \sum\limits_{j=1}^{p} \left( 
			(\hbeta^{2}_{j})^{2} - (\beta^{2}_{*,j})^{2} \right)
			\\ &+ \alpha \lambda_{s} \sum\limits_{j=1}^{p} \left(\vert\hbeta^{2}_{j}\vert - 
			\vert \beta^{2}_{*,j}\vert \right)
			\\ &+ \lambda_{d} \sum\limits_{j=1}^{p}\vert \hbeta^{1}_{j}\vert \vert 
			\hbeta^{2}_{j}\vert.
			\end{align*}
			Hence, the definition of $\bbet_{*}$ implies that
			\begin{align*}
			O(\mathbf{y}, \mathbf{X}, \bbet) - O(\mathbf{y}, \mathbf{X}, \bbet_{*}) &= 
			\frac{s_{R}}{2}\sum\limits_{j\in J} \frac{\text{soft}(r_{j}, \alpha 
				\lambda_{s})}{s_{R}+\lambda_{d}}^{2} - \frac{\text{soft}(r_{j}, \alpha 
				\lambda_{s})}{s_{R}}^{2}
			\\ & + \sum\limits_{j\in J} r_{j}\frac{\text{soft}(r_{j}, \alpha 
				\lambda_{s})}{s_{R}} - r_{j}\frac{\text{soft}(r_{j}, \alpha 
				\lambda_{s})}{s_{R}+\lambda_{d}}
			\\ &+ \alpha \lambda_{s} \frac{s_{R}}{2}\sum\limits_{j\in J} 
			\left(\left\vert\frac{\text{soft}(r_{j}, \alpha 
				\lambda_{s})}{s_{R}+\lambda_{d}}\right\vert - \left\vert 
			\frac{\text{soft}(r_{j}, \alpha \lambda_{s})}{s_{R}}\right\vert \right)
			\\&+ \frac{s_{R}}{2}\sum\limits_{j\in J} \frac{\text{soft}(r_{j}, \alpha 
				\lambda_{s})}{s_{R}+\lambda_{d}}^{2} - \sum\limits_{j\in J} 
			r_{j}\frac{\text{soft}(r_{j}, \alpha \lambda_{s})}{s_{R}+\lambda_{d}}
			\\ &+ \alpha\lambda_{s} \sum\limits_{j\in J} \left\vert\frac{\text{soft}(r_{j}, 
				\alpha \lambda_{s})}{s_{R}+\lambda_{d}}\right\vert + \lambda_{d} 
			\sum\limits_{j\in J}\left(\frac{\text{soft}(r_{j}, \alpha 
				\lambda_{s})}{s_{R}+\lambda_{d}}\right)^{2}.
			\end{align*}
			Straightforward calculations show that
			$$
			O(\mathbf{y}, \mathbf{X}, \hbbet) - O(\mathbf{y}, \mathbf{X}, \bbet_{*}) = 
			\frac{s_{R} (\lambda_{d}^{2} - s_{R}^{2})}{2}\sum\limits_{j\in J}\frac{(\vert 
				r_{j}\vert - \alpha\lambda_{s})^{2}}{(s_{R}(s_{R}+\lambda_{d}))^{2}}.
			$$
			By definition of $\hbbet$, $O(\mathbf{y}, \mathbf{X}, \hbbet) - O(\mathbf{y}, 
			\mathbf{X}, \bbet_{*}) \leq 0 $, and hence since $\vert r_{j_0}\vert > \alpha \lambda_{s}$, $\lambda_{d} \leq s_{R}$, a 
			contradiction. Thus if $\lambda_{d}>s_{R}$, only one of $\hbeta^{1}_{j_0}$ and $\hbeta^{2}_{j_0}$ is zero
			and the non-zero one has to be equal to $$\frac{\text{soft}(r_{j_0}, \alpha 
				\lambda_{s})}{s_{R}},$$
			which proves 2(c).
		\end{proof}
		
		\begin{proof}[Proof of Proposition 2]
			Let $s_{R}=1+(1-\alpha)\lambda_{S}$.
			We first prove 1. We assume that $\hbeta^{1}_{2}=\hbeta^{2}_{1}=0$, the other 
			case is similar. Then it is easy to show that
			$$
			\hbeta^{1}_{1}=T_1, \quad \hbeta^{2}_{2}=T_2.
			$$
			Since
			$$
			\hbeta^{1}_{2} = \frac{\text{soft}(r_2 - \rho T_1, \alpha \lambda_{S} + 
				\lambda_{D} \vert T_2\vert)}{s_{R}}, \quad \hbeta^{2}_{1} = 
			\frac{\text{soft}(r_1 - \rho T_2, \alpha \lambda_{S} + \lambda_{D} \vert 
				T_1\vert)}{s_{R}},
			$$
			it must be that
			\begin{align*}
			\vert r_2 - \rho T_1 \vert \leq \alpha \lambda_{s} + \lambda_{d} \vert T_2\vert, 
			\quad \vert r_1 - \rho T_2 \vert \leq \alpha \lambda_{s} + \lambda_{d} \vert 
			T_1\vert,
			\end{align*}
			from which the result follows immediately.
			
			To prove 2, assume $\hbeta^{1}_{1}=\hbeta^{2}_{1}=0$, the other case is proven 
			similarly. Then
			$$
			\hbeta^{1}_{2} = \frac{\text{soft}(r_2, \alpha \lambda_{s} + \lambda_{d} \vert 
				\hbeta^{2}_{2}\vert)}{s_{R}}, \quad \hbeta^{2}_{2} = \frac{\text{soft}(r_2, 
				\alpha \lambda_{s} + \lambda_{d} \vert \hbeta^{1}_{2}\vert)}{s_{R}}.
			$$
			Proceeding as in the proof of Proposition 1, the 
			result follows.
			
			Now we prove 3. It is easy to show that
			\begin{align*}
			&\hbeta^{1}_{1}= \text{soft}(r_1 - \rho \hbeta^{1}_{2}, \lambda_{d} \vert 
			\hbeta^{2}_{1}\vert)= r_1 - \rho \hbeta^{1}_{2}- \sign(r_1 - \rho 
			\hbeta^{1}_{2})\lambda_{d} \vert \hbeta^{2}_{1}\vert
			\\&\hbeta^{2}_{1}= \text{soft}(r_1 - \rho \hbeta^{2}_{2}, \lambda_{d} \vert 
			\hbeta^{1}_{1}\vert) = r_1 - \rho \hbeta^{2}_{2}- \sign(r_1 - \rho 
			\hbeta^{2}_{2})\lambda_{d} \vert \hbeta^{1}_{1}\vert
			\\&\hbeta^{1}_{2}= \text{soft}(r_2 - \rho \hbeta^{1}_{1}, \lambda_{d} \vert 
			\hbeta^{2}_{2}\vert) = r_2 - \rho \hbeta^{1}_{1}- \sign(r_2 - \rho 
			\hbeta^{1}_{1})\lambda_{d} \vert \hbeta^{2}_{2}\vert
			\\& \hbeta^{2}_{2}= \text{soft}(r_2 - \rho \hbeta^{2}_{1}, \lambda_{d} \vert 
			\hbeta^{1}_{2}\vert)=r_2 - \rho \hbeta^{2}_{1}- \sign(r_2 - \rho 
			\hbeta^{2}_{1})\lambda_{d} \vert \hbeta^{1}_{2}\vert.
			\end{align*}
			Hence
			\begin{equation*}
			\begin{pmatrix}
			\hbeta^{1}_{1}\\
			\hbeta^{2}_{1}\\
			\hbeta^{2}_{2}\\
			\hbeta^{1}_{2}
			\end{pmatrix}
			=
			\begin{pmatrix}
			r_{1}\\
			r_{1}\\
			r_{2}\\
			r_{2}
			\end{pmatrix}
			+
			\begin{pmatrix}
			0 &-\lambda_{d} &0& -\rho \\
			-\lambda_{d}& 0 &-\rho& 0 \\
			0 & -\rho & 0 &-\lambda_{d} \\
			-\rho & 0 & -\lambda_{d} & 0
			\end{pmatrix}
			\begin{pmatrix}
			\hbeta^{1}_{1}\\
			\hbeta^{2}_{1}\\
			\hbeta^{2}_{2}\\
			\hbeta^{1}_{2}
			\end{pmatrix}
			\end{equation*}
			and
			\begin{equation*}
			\begin{pmatrix}
			1 &\lambda_{d} &0& \rho \\
			\lambda_{d}& 1 &\rho& 0 \\
			0 & \rho & 1 &\lambda_{d} \\
			\rho & 0 & \lambda_{d} & 1
			\end{pmatrix}
			\begin{pmatrix}
			\hbeta^{1}_{1}\\
			\hbeta^{2}_{1}\\
			\hbeta^{2}_{2}\\
			\hbeta^{1}_{2}
			\end{pmatrix}
			=\begin{pmatrix}
			r_{1}\\
			r_{1}\\
			r_{2}\\
			r_{2}
			\end{pmatrix}.
			\end{equation*}
			The eigenvalues of the matrix on the left hand side are 
			$\pm\lambda_{d}\pm\rho+1$. Hence if $\lambda_{d}<1-\rho$ the linear system has a 
			unique solution.
		\end{proof}

		\begin{proof}[Proof of Theorem 1]
			Let $\bbet^{0}$ for the matrix in $\mathbb{R}^{p\times G}$ with columns
			equal to $\bbet_0$. Let $\boldsymbol{\varepsilon}=(\varepsilon_1,\dots,\varepsilon_n)^{\prime}$.
			Let $\mathbf{E}\in\mathbb{R}^{n\times G}$ be the matrix with columns equal to $\boldsymbol{\varepsilon}$.
			Then the linear model can be written as
			$$
			\mathbf{Y}=\mathbf{X}\bbet^{0}+\mathbf{E}.
			$$
			Let
			$$
			q(\bbet) = \left( \Vert  \vert\bbet\vert^{\prime} \vert\bbet\vert\Vert_1 - 
			\Vert\bbet\Vert_{F}^{2} \right)
			$$
			and
			$$
			P(\bbet)=\left( \frac{(1-\alpha)}{2}\Vert \bbet\Vert _{F}^{2}+\alpha \Vert \bbet \Vert_{1}\right).
			$$
			Let $\hbbet$ be any solution to SplitReg. Then
			\begin{align*}
			\frac{1}{2n}\Vert \mathbf{Y} - \mathbf{X} 
			\hbbet\Vert_{F}^{2} + \lambda_{s} P(\hbbet)+ \frac{\lambda_{d}}{2} 
			q(\hbbet) &\leq
			\frac{1}{2n}\Vert \mathbf{Y} - \mathbf{X} 
			\bbet^{0}\Vert_{F}^{2} + \lambda_{s} P(\bbet^{0})+ \frac{\lambda_{d}}{2} 
			q(\bbet^{0})
			\\ &= \frac{1}{2n}\Vert \mathbf{E}\Vert_{F}^{2} + \lambda_{s}P(\bbet^{0})+ \frac{\lambda_{d}}{2} 
			q(\bbet^{0}).
			\end{align*}
			Hence
			\begin{align*}
			\frac{1}{2n}\Vert \mathbf{X}\hbbet -  \mathbf{X}\bbet^{0}
			\Vert_{F}^{2} + \lambda_{s} P(\hbbet)+ \frac{\lambda_{d}}{2} 
			q(\hbbet) &\leq \frac{1}{n}\Tr\left(\mathbf{E}^{\prime} \mathbf{X}\left( \hbbet - \bbet^{0}\right) \right) 
			+ \lambda_{s} P(\bbet^{0})+ \frac{\lambda_{d}}{2} 
			q(\bbet^{0})
			\\ &\leq \frac{1}{n} \Vert \mathbf{X}^{\prime} \mathbf{E} \Vert_{\infty}\Vert \hbbet - \bbet^{0}\Vert_{1} + \lambda_{s} P(\bbet^{0})+ \frac{\lambda_{d}}{2} 
			q(\bbet^{0}),
			\end{align*}
			where $\Vert \cdot \Vert_{\infty}$ is the coordinate-wise maximum absolute value and $\Vert \cdot \Vert_{1}$ is the sum of the absolute values.
			It follows that
			\begin{align*}
			\frac{1}{2n}\Vert \mathbf{X}\hbbet -  \mathbf{X}\bbet^{0}
			\Vert_{F}^{2} &\leq \frac{1}{n} \Vert \mathbf{X}^{\prime} \mathbf{E} \Vert_{\infty} \left(\Vert \hbbet\Vert_{1} +\Vert \bbet^{0}\Vert_{1}\right)+
			\lambda_{s}\left(  P(\bbet^{0})- P(\hbbet)   \right)+\frac{\lambda_{d}}{2}q(\bbet^{0})
			\\&= \frac{1}{n} \max_{1\leq j \leq p }\vert \boldsymbol{\varepsilon}^{\prime}\mathbf{x}^{j} \vert \left(\Vert \hbbet\Vert_{1} +\Vert \bbet^{0}\Vert_{1}\right)+
			\lambda_{s}\left(P(\bbet^{0})- P(\hbbet) \right)+\frac{\lambda_{d}}{2}q(\bbet^{0}).
			\end{align*}
			Since $U_j=\boldsymbol{\varepsilon}^{\prime}\mathbf{x}^{j} (\sqrt{n} \sigma)^{-1}$ is standard normal
			$$
			P\left( \max_{1\leq j \leq p} \vert U_j \vert > \sqrt{t^2+2\log(p)}\right)\leq p \exp\left( -\frac{t^2+2\log p}{2} \right)=2\exp(-t^2/2).
			$$
			Thus with probability at least $1-2\exp(-t^2/2)$
			\begin{align*}
			\frac{1}{2n}\Vert \mathbf{X}\hbbet -  \mathbf{X}\bbet^{0}
			\Vert_{F}^{2} &\leq \frac{1}{n} \max_{1\leq j \leq p }\vert \boldsymbol{\varepsilon}^{\prime}\mathbf{x}^{j} \vert \left(\Vert \hbbet\Vert_{1} +\Vert \bbet^{0}\Vert_{1}\right)+
			\lambda_{s}\left( P(\bbet^{0})- P(\hbbet)  \right)+\frac{\lambda_{d}}{2}q(\bbet^{0})
			\\ &\leq \sigma \sqrt{\frac{t^2+2\log(p)}{n}} \Vert \bbet^{0}\Vert_{1}+\lambda_{s}P(\bbet^{0})+\frac{\lambda_{d}}{2}q(\bbet^{0})
			\\ &+ \left(\sigma \sqrt{\frac{t^2+2\log(p)}{n}} -\alpha \lambda_{s}\right)\Vert \hbbet\Vert_{1} - \frac{\lambda_{s}(1-\alpha)}{2} \Vert \hbbet\Vert_{F}^{2}
			\\ &\leq \alpha \lambda_{s}\Vert \bbet^{0}\Vert_{1} + \lambda_{s}P(\bbet^{0}) + \frac{\lambda_{d}}{2}q(\bbet^{0})
			\\ & = 2\alpha \lambda_{s} \Vert \bbet^0\Vert_{1} + \lambda_{s}\frac{(1-\alpha)}{2}\Vert \bbet^{0}\Vert _{F}^{2}+ \frac{\lambda_{d}}{2}q(\bbet^{0}).
			\end{align*}
			Since $q(\bbet^{0})=(G-1)\Vert \bbet^{0}\Vert_{F}^2$, we have 
			$$
			\frac{1}{2n}\Vert \mathbf{X}\hbbet -  \mathbf{X}\bbet^{0}
			\Vert_{F}^{2} \leq 2\alpha \lambda_{s} \Vert \bbet^0\Vert_{1} + \lambda_{s}\frac{(1-\alpha)}{2}\Vert \bbet^{0}\Vert _{F}^{2}+ \frac{\lambda_{d} (G-1)}{2} \Vert \bbet^{0} \Vert_{F}^2.
			$$
			Hence
			\begin{equation}
			\frac{1}{2n} \sum\limits_{g=1}^{G}\Vert \mathbf{X}\hbbet^{g} -  \mathbf{X}\bbet_{0}
			\Vert_{2}^{2}\leq 2\alpha \lambda_{s} G\Vert \bbet_0\Vert_{1} + \lambda_{s}G\frac{(1-\alpha)}{2}\Vert \bbet_{0}\Vert_{2}^{2}+ \frac{\lambda_{d} G(G-1)}{2} \Vert \bbet_{0} \Vert_{2}^2.
			\label{eq:bound_sum}
			\end{equation}
			
			Note that
			\begin{align*}
			\frac{1}{2n} \left\Vert \frac{1}{G}\sum\limits_{g=1}^{G}\mathbf{X}\hbbet^{g} -  \mathbf{X}\bbet_{0}
			\right\Vert_{2}^{2}&=\frac{1}{2nG^2} \left\Vert \sum\limits_{g=1}^{G}\left(\mathbf{X}\hbbet^{g} -  \mathbf{X}\bbet_{0}\right)
			\right\Vert_{2}^{2}\leq \frac{1}{2nG^2} \left( \sum\limits_{g=1}^{G}\left\Vert\mathbf{X}\hbbet^{g} -  \mathbf{X}\bbet_{0}\right\Vert_{2} \right)^2
			\\ &\leq \frac{1}{2nG^2} \left( \left(\sum\limits_{g=1}^{G}\left\Vert\mathbf{X}\hbbet^{g} -  \mathbf{X}\bbet_{0}\right\Vert_{2}^{2} \right)^{1/2} \sqrt{G} \right)^2
			\\ &= \frac{1}{2nG} \sum\limits_{g=1}^{G}\left\Vert\mathbf{X}\hbbet^{g} -  \mathbf{X}\bbet_{0}\right\Vert_{2}^{2}
			\\ & \leq 2\alpha \lambda_{s} \Vert \bbet_0\Vert_{1} + \lambda_{s}\frac{(1-\alpha)}{2}\Vert \bbet_{0}\Vert _{2}^{2}+ \frac{\lambda_{d} (G-1)}{2} \Vert \bbet_0 \Vert_{2}^2,
			\end{align*}
			where we have used \eqref{eq:bound_sum} and the inequality $\Vert \mathbf{v}\Vert_{1}\leq\Vert \mathbf{v}\Vert_{2}\sqrt{G}$ for $\mathbf{v}\in\mathbb{R}^{G}$. The Theorem is proven.

		\end{proof}

		\section*{Appendix B: Full results of the simulation study}
		
		The maximum standard error reported in the caption of each table does not include cases where the average prediction mean squared errors was greater than 3.
		\input{sim_results.txt}

	\end{document}

%% file: numgroups_results.txt
\begin{table}[ht]
\caption{MSPEs, overlap and average computation time in seconds for different values of the number of models (rows) and proportion of active variables $\zeta$ (columns) for SNR=10.} 
\centering
\begin{tabular}{rrrrrrrr}
   \hline & \multicolumn{2}{c}{$\mathbf{\boldsymbol{\zeta}=0.1}$} & \multicolumn{2}{c}{$\mathbf{\boldsymbol{\zeta}=0.2}$} & \multicolumn{2}{c}{$\mathbf{\boldsymbol{\zeta}=0.3}$} & \\ \cmidrule(lr){2-3} \cmidrule(lr){4-5} \cmidrule(lr){6-7} $\boldsymbol{G}$ & \textbf{MSPE} & \textbf{OVP} & \textbf{MSPE} & \textbf{OVP} & \textbf{MSPE} & \textbf{OVP} & \textbf{Time} \\ \cmidrule(lr){1-1} \cmidrule(lr){2-3} \cmidrule(lr){4-5} \cmidrule(lr){6-7} \cmidrule(lr){8-8} 2 & 1.27 & 0.64 & 1.23 & 0.66 & 1.23 & 0.63 & 8.61 \\ 
  5 & 1.22 & 0.44 & 1.18 & 0.43 & 1.17 & 0.37 & 19.25 \\ 
  7 & 1.22 & 0.42 & 1.17 & 0.38 & 1.17 & 0.36 & 25.23 \\ 
  10 & 1.21 & 0.38 & 1.16 & 0.36 & 1.16 & 0.33 & 33.97 \\ 
  15 & 1.19 & 0.35 & 1.15 & 0.36 & 1.15 & 0.32 & 48.98 \\ 
  20 & 1.20 & 0.35 & 1.15 & 0.34 & 1.15 & 0.30 & 64.05 \\ 
  30 & 1.19 & 0.31 & 1.15 & 0.33 & 1.15 & 0.29 & 95.69 \\ 
   \hline
\end{tabular}
\label{tab:numgroups}
\end{table}

%% file: sim_results.txt
\begin{table}[ht]
\caption{Mean MSPEs, recalls and precisions for Scenario 1 with SNR = 3, $\rho=0.2$, $n=100$, $p=1000$. MSPEs maximum standard error is 0.03.} 
\centering
\resizebox{\textwidth}{!}{
\begin{tabular}{lrrrrrrrrrrrr}
   \hline & \multicolumn{3}{c}{$\mathbf{\boldsymbol{\zeta}=0.05}$} & \multicolumn{3}{c}{$\mathbf{\boldsymbol{\zeta}=0.1}$} & \multicolumn{3}{c}{$\mathbf{\boldsymbol{\zeta}=0.2}$} & \multicolumn{3}{c}{$\mathbf{\boldsymbol{\zeta}=0.4}$} \\ \cmidrule(lr){2-4} \cmidrule(lr){5-7} \cmidrule(lr){8-10} \cmidrule(lr){11-13} \textbf{Method} & \textbf{MSPE} & \textbf{RC} & \textbf{PR} & \textbf{MSPE} & \textbf{RC} & \textbf{PR} & \textbf{MSPE} & \textbf{RC} & \textbf{PR} & \textbf{MSPE} & \textbf{RC} & \textbf{PR} \\ \cmidrule(lr){1-1} \cmidrule(lr){2-4} \cmidrule(lr){5-7} \cmidrule(lr){8-10} \cmidrule(lr){11-13}Lasso & 1.74 & 0.15 & 0.13 & 1.59 & 0.10 & 0.16 & 1.46 & 0.08 & 0.25 & 1.44 & 0.07 & 0.44 \\ 
  Elastic Net & 1.72 & 0.16 & 0.13 & 1.57 & 0.11 & 0.16 & 1.45 & 0.08 & 0.25 & 1.41 & 0.08 & 0.43 \\ 
  SplitReg-Lasso & 1.58 & 0.41 & 0.09 & 1.38 & 0.37 & 0.14 & 1.28 & 0.34 & 0.23 & 1.22 & 0.32 & 0.42 \\ 
  SplitReg-EN & 1.56 & 0.42 & 0.10 & 1.38 & 0.37 & 0.14 & 1.27 & 0.36 & 0.23 & 1.21 & 0.35 & 0.42 \\ 
  SparseNet & 1.76 & 0.16 & 0.14 & 1.63 & 0.11 & 0.17 & 1.50 & 0.09 & 0.26 & 1.46 & 0.08 & 0.45 \\ 
  Relaxed & 1.80 & 0.12 & 0.15 & 1.66 & 0.07 & 0.17 & 1.54 & 0.06 & 0.25 & 1.51 & 0.06 & 0.44 \\ 
  Stepwise & 2.83 & 0.02 & 0.15 & 2.59 & 0.01 & 0.18 & 2.46 & 0.01 & 0.26 & 2.41 & 0.01 & 0.44 \\ 
  RF & 1.94 & $ - $ & $ - $ & 1.82 & $ - $ & $ - $ & 1.77 & $ - $ & $ - $ & 1.72 & $ - $ & $ - $ \\ 
  RGLM & 1.50 & 0.77 & 0.06 & 1.32 & 0.74 & 0.11 & 1.23 & 0.71 & 0.21 & 1.17 & 0.70 & 0.40 \\ 
  SIS-SCAD & 2.65 & 0.05 & 0.13 & 2.41 & 0.03 & 0.15 & 2.24 & 0.03 & 0.24 & 2.17 & 0.02 & 0.43 \\ 
   \hline
\end{tabular}}
\end{table}
\begin{table}[ht]
\caption{Mean MSPEs, recalls and precisions for Scenario 1 with SNR = 5, $\rho=0.2$, $n=100$, $p=1000$. MSPEs maximum standard error is 0.03.} 
\centering
\resizebox{\textwidth}{!}{\begin{tabular}{lrrrrrrrrrrrr}
   \hline & \multicolumn{3}{c}{$\mathbf{\boldsymbol{\zeta}=0.05}$} & \multicolumn{3}{c}{$\mathbf{\boldsymbol{\zeta}=0.1}$} & \multicolumn{3}{c}{$\mathbf{\boldsymbol{\zeta}=0.2}$} & \multicolumn{3}{c}{$\mathbf{\boldsymbol{\zeta}=0.4}$} \\ \cmidrule(lr){2-4} \cmidrule(lr){5-7} \cmidrule(lr){8-10} \cmidrule(lr){11-13} \textbf{Method} & \textbf{MSPE} & \textbf{RC} & \textbf{PR} & \textbf{MSPE} & \textbf{RC} & \textbf{PR} & \textbf{MSPE} & \textbf{RC} & \textbf{PR} & \textbf{MSPE} & \textbf{RC} & \textbf{PR} \\ \cmidrule(lr){1-1} \cmidrule(lr){2-4} \cmidrule(lr){5-7} \cmidrule(lr){8-10} \cmidrule(lr){11-13}Lasso & 2.09 & 0.19 & 0.15 & 1.85 & 0.13 & 0.18 & 1.66 & 0.10 & 0.26 & 1.59 & 0.09 & 0.44 \\ 
  Elastic Net & 2.07 & 0.20 & 0.15 & 1.84 & 0.13 & 0.18 & 1.64 & 0.10 & 0.26 & 1.56 & 0.09 & 0.44 \\ 
  SplitReg-Lasso & 1.87 & 0.49 & 0.10 & 1.56 & 0.45 & 0.14 & 1.37 & 0.44 & 0.23 & 1.27 & 0.42 & 0.42 \\ 
  SplitReg-EN & 1.85 & 0.48 & 0.10 & 1.55 & 0.45 & 0.14 & 1.37 & 0.44 & 0.23 & 1.27 & 0.44 & 0.42 \\ 
  SparseNet & 2.11 & 0.20 & 0.16 & 1.89 & 0.14 & 0.19 & 1.67 & 0.12 & 0.28 & 1.59 & 0.10 & 0.47 \\ 
  Relaxed & 2.19 & 0.16 & 0.16 & 1.96 & 0.10 & 0.18 & 1.74 & 0.08 & 0.27 & 1.66 & 0.07 & 0.45 \\ 
  Stepwise &$>3$& 0.03 & 0.17 &$>3$& 0.02 & 0.18 &$>3$& 0.01 & 0.26 & 2.94 & 0.01 & 0.44 \\ 
  RF & 2.64 & $ - $ & $ - $ & 2.46 & $ - $ & $ - $ & 2.36 & $ - $ & $ - $ & 2.29 & $ - $ & $ - $ \\ 
  RGLM & 1.80 & 0.78 & 0.06 & 1.54 & 0.75 & 0.11 & 1.39 & 0.72 & 0.21 & 1.31 & 0.71 & 0.40 \\ 
  SIS-SCAD &$>3$& 0.06 & 0.15 & 2.81 & 0.03 & 0.17 & 2.57 & 0.03 & 0.25 & 2.45 & 0.02 & 0.44 \\ 
   \hline
\end{tabular}}
\end{table}
\begin{table}[ht]
\caption{Mean MSPEs, recalls and precisions for Scenario 1 with SNR = 10, $\rho=0.2$, $n=100$, $p=1000$. MSPEs maximum standard error is 0.03.} 
\centering
\resizebox{\textwidth}{!}{\begin{tabular}{lrrrrrrrrrrrr}
   \hline & \multicolumn{3}{c}{$\mathbf{\boldsymbol{\zeta}=0.05}$} & \multicolumn{3}{c}{$\mathbf{\boldsymbol{\zeta}=0.1}$} & \multicolumn{3}{c}{$\mathbf{\boldsymbol{\zeta}=0.2}$} & \multicolumn{3}{c}{$\mathbf{\boldsymbol{\zeta}=0.4}$} \\ \cmidrule(lr){2-4} \cmidrule(lr){5-7} \cmidrule(lr){8-10} \cmidrule(lr){11-13} \textbf{Method} & \textbf{MSPE} & \textbf{RC} & \textbf{PR} & \textbf{MSPE} & \textbf{RC} & \textbf{PR} & \textbf{MSPE} & \textbf{RC} & \textbf{PR} & \textbf{MSPE} & \textbf{RC} & \textbf{PR} \\ \cmidrule(lr){1-1} \cmidrule(lr){2-4} \cmidrule(lr){5-7} \cmidrule(lr){8-10} \cmidrule(lr){11-13}Lasso & 2.91 & 0.26 & 0.17 & 2.47 & 0.16 & 0.20 & 2.09 & 0.12 & 0.27 & 1.94 & 0.11 & 0.46 \\ 
  Elastic Net & 2.88 & 0.26 & 0.17 & 2.43 & 0.17 & 0.19 & 2.07 & 0.13 & 0.27 & 1.90 & 0.11 & 0.45 \\ 
  SplitReg-Lasso & 2.56 & 0.57 & 0.11 & 1.96 & 0.57 & 0.14 & 1.60 & 0.57 & 0.23 & 1.40 & 0.58 & 0.42 \\ 
  SplitReg-EN & 2.56 & 0.56 & 0.11 & 1.98 & 0.56 & 0.14 & 1.60 & 0.57 & 0.23 & 1.40 & 0.58 & 0.42 \\ 
  SparseNet & 2.88 & 0.29 & 0.18 & 2.45 & 0.19 & 0.21 & 2.06 & 0.15 & 0.29 & 1.89 & 0.13 & 0.49 \\ 
  Relaxed &$>3$& 0.21 & 0.18 & 2.60 & 0.13 & 0.20 & 2.17 & 0.10 & 0.27 & 2.00 & 0.09 & 0.46 \\ 
  Stepwise &$>3$& 0.04 & 0.18 &$>3$& 0.03 & 0.19 &$>3$& 0.02 & 0.26 &$>3$& 0.02 & 0.45 \\ 
  RF &$>3$& $ - $ & $ - $ &$>3$& $ - $ & $ - $ &$>3$& $ - $ & $ - $ &$>3$& $ - $ & $ - $ \\ 
  RGLM & 2.60 & 0.79 & 0.06 & 2.14 & 0.76 & 0.11 & 1.85 & 0.73 & 0.21 & 1.69 & 0.72 & 0.40 \\ 
  SIS-SCAD &$>3$& 0.08 & 0.18 &$>3$& 0.04 & 0.19 &$>3$& 0.03 & 0.26 &$>3$& 0.02 & 0.44 \\ 
   \hline
\end{tabular}}
\end{table}
\begin{table}[ht]
\caption{Mean MSPEs, recalls and precisions for Scenario 1 with SNR = 3, $\rho=0.5$, $n=100$, $p=1000$. MSPEs maximum standard error is 0.02.} 
\centering
\resizebox{\textwidth}{!}{\begin{tabular}{lrrrrrrrrrrrr}
   \hline & \multicolumn{3}{c}{$\mathbf{\boldsymbol{\zeta}=0.05}$} & \multicolumn{3}{c}{$\mathbf{\boldsymbol{\zeta}=0.1}$} & \multicolumn{3}{c}{$\mathbf{\boldsymbol{\zeta}=0.2}$} & \multicolumn{3}{c}{$\mathbf{\boldsymbol{\zeta}=0.4}$} \\ \cmidrule(lr){2-4} \cmidrule(lr){5-7} \cmidrule(lr){8-10} \cmidrule(lr){11-13} \textbf{Method} & \textbf{MSPE} & \textbf{RC} & \textbf{PR} & \textbf{MSPE} & \textbf{RC} & \textbf{PR} & \textbf{MSPE} & \textbf{RC} & \textbf{PR} & \textbf{MSPE} & \textbf{RC} & \textbf{PR} \\ \cmidrule(lr){1-1} \cmidrule(lr){2-4} \cmidrule(lr){5-7} \cmidrule(lr){8-10} \cmidrule(lr){11-13}Lasso & 1.31 & 0.08 & 0.10 & 1.25 & 0.06 & 0.14 & 1.21 & 0.05 & 0.22 & 1.21 & 0.05 & 0.43 \\ 
  Elastic Net & 1.30 & 0.08 & 0.10 & 1.23 & 0.06 & 0.14 & 1.21 & 0.05 & 0.22 & 1.20 & 0.05 & 0.43 \\ 
  SplitReg-Lasso & 1.22 & 0.25 & 0.08 & 1.15 & 0.22 & 0.13 & 1.13 & 0.19 & 0.22 & 1.12 & 0.19 & 0.42 \\ 
  SplitReg-EN & 1.22 & 0.25 & 0.08 & 1.15 & 0.22 & 0.12 & 1.14 & 0.20 & 0.22 & 1.11 & 0.20 & 0.42 \\ 
  SparseNet & 1.35 & 0.08 & 0.11 & 1.28 & 0.06 & 0.15 & 1.24 & 0.05 & 0.24 & 1.24 & 0.04 & 0.45 \\ 
  Relaxed & 1.37 & 0.06 & 0.10 & 1.31 & 0.04 & 0.14 & 1.26 & 0.04 & 0.23 & 1.26 & 0.03 & 0.43 \\ 
  Stepwise & 1.84 & 0.01 & 0.11 & 1.77 & 0.01 & 0.16 & 1.72 & 0.01 & 0.22 & 1.69 & 0.01 & 0.43 \\ 
  RF & 1.30 & $ - $ & $ - $ & 1.23 & $ - $ & $ - $ & 1.22 & $ - $ & $ - $ & 1.20 & $ - $ & $ - $ \\ 
  RGLM & 1.22 & 0.73 & 0.05 & 1.15 & 0.70 & 0.10 & 1.13 & 0.69 & 0.20 & 1.11 & 0.68 & 0.40 \\ 
  SIS-SCAD & 2.23 & 0.03 & 0.08 & 2.12 & 0.02 & 0.11 & 2.03 & 0.02 & 0.21 & 2.06 & 0.02 & 0.40 \\ 
   \hline
\end{tabular}}
\end{table}
\begin{table}[ht]
\caption{Mean MSPEs, recalls and precisions for Scenario 1 with SNR = 5, $\rho=0.5$, $n=100$, $p=1000$. MSPEs maximum standard error is 0.02.} 
\centering
\resizebox{\textwidth}{!}{\begin{tabular}{lrrrrrrrrrrrr}
   \hline & \multicolumn{3}{c}{$\mathbf{\boldsymbol{\zeta}=0.05}$} & \multicolumn{3}{c}{$\mathbf{\boldsymbol{\zeta}=0.1}$} & \multicolumn{3}{c}{$\mathbf{\boldsymbol{\zeta}=0.2}$} & \multicolumn{3}{c}{$\mathbf{\boldsymbol{\zeta}=0.4}$} \\ \cmidrule(lr){2-4} \cmidrule(lr){5-7} \cmidrule(lr){8-10} \cmidrule(lr){11-13} \textbf{Method} & \textbf{MSPE} & \textbf{RC} & \textbf{PR} & \textbf{MSPE} & \textbf{RC} & \textbf{PR} & \textbf{MSPE} & \textbf{RC} & \textbf{PR} & \textbf{MSPE} & \textbf{RC} & \textbf{PR} \\ \cmidrule(lr){1-1} \cmidrule(lr){2-4} \cmidrule(lr){5-7} \cmidrule(lr){8-10} \cmidrule(lr){11-13}Lasso & 1.43 & 0.10 & 0.11 & 1.34 & 0.07 & 0.15 & 1.28 & 0.06 & 0.23 & 1.27 & 0.06 & 0.43 \\ 
  Elastic Net & 1.42 & 0.11 & 0.11 & 1.32 & 0.08 & 0.15 & 1.28 & 0.06 & 0.23 & 1.25 & 0.06 & 0.43 \\ 
  SplitReg-Lasso & 1.31 & 0.31 & 0.09 & 1.20 & 0.28 & 0.13 & 1.17 & 0.24 & 0.22 & 1.14 & 0.24 & 0.42 \\ 
  SplitReg-EN & 1.31 & 0.32 & 0.08 & 1.20 & 0.28 & 0.13 & 1.16 & 0.26 & 0.22 & 1.13 & 0.25 & 0.42 \\ 
  SparseNet & 1.46 & 0.11 & 0.13 & 1.37 & 0.08 & 0.17 & 1.31 & 0.07 & 0.26 & 1.29 & 0.06 & 0.46 \\ 
  Relaxed & 1.48 & 0.08 & 0.11 & 1.42 & 0.05 & 0.15 & 1.34 & 0.04 & 0.24 & 1.33 & 0.04 & 0.43 \\ 
  Stepwise & 2.15 & 0.02 & 0.13 & 2.03 & 0.01 & 0.16 & 1.95 & 0.01 & 0.23 & 1.90 & 0.01 & 0.42 \\ 
  RF & 1.52 & $ - $ & $ - $ & 1.43 & $ - $ & $ - $ & 1.41 & $ - $ & $ - $ & 1.37 & $ - $ & $ - $ \\ 
  RGLM & 1.29 & 0.75 & 0.05 & 1.18 & 0.72 & 0.11 & 1.14 & 0.70 & 0.20 & 1.11 & 0.69 & 0.40 \\ 
  SIS-SCAD & 2.32 & 0.04 & 0.09 & 2.20 & 0.03 & 0.12 & 2.09 & 0.02 & 0.22 & 2.07 & 0.02 & 0.42 \\ 
   \hline
\end{tabular}}
\end{table}
\begin{table}[ht]
\caption{Mean MSPEs, recalls and precisions for Scenario 1 with SNR = 10, $\rho=0.5$, $n=100$, $p=1000$. MSPEs maximum standard error is 0.03.} 
\centering
\resizebox{\textwidth}{!}{\begin{tabular}{lrrrrrrrrrrrr}
   \hline & \multicolumn{3}{c}{$\mathbf{\boldsymbol{\zeta}=0.05}$} & \multicolumn{3}{c}{$\mathbf{\boldsymbol{\zeta}=0.1}$} & \multicolumn{3}{c}{$\mathbf{\boldsymbol{\zeta}=0.2}$} & \multicolumn{3}{c}{$\mathbf{\boldsymbol{\zeta}=0.4}$} \\ \cmidrule(lr){2-4} \cmidrule(lr){5-7} \cmidrule(lr){8-10} \cmidrule(lr){11-13} \textbf{Method} & \textbf{MSPE} & \textbf{RC} & \textbf{PR} & \textbf{MSPE} & \textbf{RC} & \textbf{PR} & \textbf{MSPE} & \textbf{RC} & \textbf{PR} & \textbf{MSPE} & \textbf{RC} & \textbf{PR} \\ \cmidrule(lr){1-1} \cmidrule(lr){2-4} \cmidrule(lr){5-7} \cmidrule(lr){8-10} \cmidrule(lr){11-13}Lasso & 1.71 & 0.15 & 0.13 & 1.55 & 0.10 & 0.16 & 1.42 & 0.08 & 0.25 & 1.39 & 0.07 & 0.44 \\ 
  Elastic Net & 1.69 & 0.16 & 0.13 & 1.53 & 0.10 & 0.16 & 1.41 & 0.08 & 0.25 & 1.37 & 0.07 & 0.43 \\ 
  SplitReg-Lasso & 1.52 & 0.42 & 0.09 & 1.33 & 0.37 & 0.14 & 1.24 & 0.33 & 0.23 & 1.18 & 0.32 & 0.42 \\ 
  SplitReg-EN & 1.52 & 0.42 & 0.09 & 1.33 & 0.38 & 0.14 & 1.23 & 0.34 & 0.23 & 1.18 & 0.33 & 0.42 \\ 
  SparseNet & 1.73 & 0.17 & 0.15 & 1.54 & 0.12 & 0.19 & 1.43 & 0.10 & 0.28 & 1.40 & 0.08 & 0.48 \\ 
  Relaxed & 1.77 & 0.12 & 0.14 & 1.62 & 0.08 & 0.16 & 1.49 & 0.06 & 0.25 & 1.45 & 0.05 & 0.44 \\ 
  Stepwise & 2.81 & 0.02 & 0.14 & 2.56 & 0.01 & 0.16 & 2.39 & 0.01 & 0.26 & 2.32 & 0.01 & 0.42 \\ 
  RF & 2.12 & $ - $ & $ - $ & 1.98 & $ - $ & $ - $ & 1.91 & $ - $ & $ - $ & 1.86 & $ - $ & $ - $ \\ 
  RGLM & 1.50 & 0.76 & 0.06 & 1.31 & 0.74 & 0.11 & 1.23 & 0.72 & 0.21 & 1.17 & 0.70 & 0.40 \\ 
  SIS-SCAD & 2.74 & 0.05 & 0.11 & 2.48 & 0.03 & 0.14 & 2.29 & 0.02 & 0.24 & 2.28 & 0.02 & 0.42 \\ 
   \hline
\end{tabular}}
\end{table}
\begin{table}[ht]
\caption{Mean MSPEs, recalls and precisions for Scenario 1 with SNR = 3, $\rho=0.8$, $n=100$, $p=1000$. MSPEs maximum standard error is 0.02.} 
\centering
\resizebox{\textwidth}{!}{\begin{tabular}{lrrrrrrrrrrrr}
   \hline & \multicolumn{3}{c}{$\mathbf{\boldsymbol{\zeta}=0.05}$} & \multicolumn{3}{c}{$\mathbf{\boldsymbol{\zeta}=0.1}$} & \multicolumn{3}{c}{$\mathbf{\boldsymbol{\zeta}=0.2}$} & \multicolumn{3}{c}{$\mathbf{\boldsymbol{\zeta}=0.4}$} \\ \cmidrule(lr){2-4} \cmidrule(lr){5-7} \cmidrule(lr){8-10} \cmidrule(lr){11-13} \textbf{Method} & \textbf{MSPE} & \textbf{RC} & \textbf{PR} & \textbf{MSPE} & \textbf{RC} & \textbf{PR} & \textbf{MSPE} & \textbf{RC} & \textbf{PR} & \textbf{MSPE} & \textbf{RC} & \textbf{PR} \\ \cmidrule(lr){1-1} \cmidrule(lr){2-4} \cmidrule(lr){5-7} \cmidrule(lr){8-10} \cmidrule(lr){11-13}Lasso & 1.14 & 0.04 & 0.07 & 1.11 & 0.03 & 0.12 & 1.12 & 0.03 & 0.21 & 1.11 & 0.03 & 0.41 \\ 
  Elastic Net & 1.14 & 0.04 & 0.07 & 1.10 & 0.04 & 0.12 & 1.11 & 0.03 & 0.21 & 1.10 & 0.03 & 0.42 \\ 
  SplitReg-Lasso & 1.10 & 0.14 & 0.07 & 1.07 & 0.12 & 0.12 & 1.08 & 0.11 & 0.21 & 1.07 & 0.11 & 0.41 \\ 
  SplitReg-EN & 1.10 & 0.15 & 0.07 & 1.07 & 0.13 & 0.12 & 1.08 & 0.12 & 0.21 & 1.07 & 0.12 & 0.41 \\ 
  SparseNet & 1.20 & 0.04 & 0.10 & 1.15 & 0.04 & 0.14 & 1.16 & 0.03 & 0.25 & 1.16 & 0.03 & 0.46 \\ 
  Relaxed & 1.20 & 0.03 & 0.08 & 1.15 & 0.02 & 0.12 & 1.15 & 0.02 & 0.22 & 1.16 & 0.02 & 0.41 \\ 
  Stepwise & 1.38 & $ - $ & 0.08 & 1.36 & $ 0.00 $ & 0.12 & 1.34 & $ 0.00 $ & 0.23 & 1.36 & $ 0.00 $ & 0.41 \\ 
  RF & 1.12 & $ - $ & $ - $ & 1.07 & $ - $ & $ - $ & 1.09 & $ - $ & $ - $ & 1.08 & $ - $ & $ - $ \\ 
  RGLM & 1.15 & 0.66 & 0.05 & 1.11 & 0.64 & 0.10 & 1.11 & 0.63 & 0.20 & 1.11 & 0.62 & 0.40 \\ 
  SIS-SCAD & 2.05 & 0.02 & 0.06 & 2.02 & 0.02 & 0.11 & 1.98 & 0.02 & 0.20 & 1.97 & 0.02 & 0.40 \\ 
   \hline
\end{tabular}}
\end{table}
\begin{table}[ht]
\caption{Mean MSPEs, recalls and precisions for Scenario 1 with SNR = 5, $\rho=0.8$, $n=100$, $p=1000$. MSPEs maximum standard error is 0.02.} 
\centering
\resizebox{\textwidth}{!}{\begin{tabular}{lrrrrrrrrrrrr}
   \hline & \multicolumn{3}{c}{$\mathbf{\boldsymbol{\zeta}=0.05}$} & \multicolumn{3}{c}{$\mathbf{\boldsymbol{\zeta}=0.1}$} & \multicolumn{3}{c}{$\mathbf{\boldsymbol{\zeta}=0.2}$} & \multicolumn{3}{c}{$\mathbf{\boldsymbol{\zeta}=0.4}$} \\ \cmidrule(lr){2-4} \cmidrule(lr){5-7} \cmidrule(lr){8-10} \cmidrule(lr){11-13} \textbf{Method} & \textbf{MSPE} & \textbf{RC} & \textbf{PR} & \textbf{MSPE} & \textbf{RC} & \textbf{PR} & \textbf{MSPE} & \textbf{RC} & \textbf{PR} & \textbf{MSPE} & \textbf{RC} & \textbf{PR} \\ \cmidrule(lr){1-1} \cmidrule(lr){2-4} \cmidrule(lr){5-7} \cmidrule(lr){8-10} \cmidrule(lr){11-13}Lasso & 1.18 & 0.05 & 0.08 & 1.14 & 0.04 & 0.12 & 1.14 & 0.04 & 0.22 & 1.13 & 0.03 & 0.41 \\ 
  Elastic Net & 1.18 & 0.05 & 0.08 & 1.13 & 0.04 & 0.13 & 1.14 & 0.04 & 0.22 & 1.13 & 0.04 & 0.42 \\ 
  SplitReg-Lasso & 1.13 & 0.18 & 0.07 & 1.08 & 0.16 & 0.12 & 1.09 & 0.15 & 0.22 & 1.08 & 0.14 & 0.42 \\ 
  SplitReg-EN & 1.13 & 0.19 & 0.07 & 1.08 & 0.16 & 0.12 & 1.09 & 0.15 & 0.22 & 1.08 & 0.14 & 0.41 \\ 
  SparseNet & 1.24 & 0.06 & 0.11 & 1.17 & 0.05 & 0.16 & 1.18 & 0.04 & 0.26 & 1.17 & 0.04 & 0.46 \\ 
  Relaxed & 1.25 & 0.03 & 0.09 & 1.20 & 0.03 & 0.12 & 1.19 & 0.03 & 0.22 & 1.19 & 0.02 & 0.42 \\ 
  Stepwise & 1.51 & 0.01 & 0.09 & 1.47 & $ 0.00 $ & 0.11 & 1.46 & $ 0.00 $ & 0.22 & 1.46 & $ 0.00 $ & 0.41 \\ 
  RF & 1.17 & $ - $ & $ - $ & 1.11 & $ - $ & $ - $ & 1.13 & $ - $ & $ - $ & 1.12 & $ - $ & $ - $ \\ 
  RGLM & 1.17 & 0.69 & 0.05 & 1.12 & 0.67 & 0.10 & 1.12 & 0.66 & 0.20 & 1.11 & 0.65 & 0.40 \\ 
  SIS-SCAD & 2.15 & 0.03 & 0.06 & 2.10 & 0.02 & 0.11 & 2.06 & 0.02 & 0.21 & 2.03 & 0.02 & 0.40 \\ 
   \hline
\end{tabular}}
\end{table}
\begin{table}[ht]
\caption{Mean MSPEs, recalls and precisions for Scenario 1 with SNR = 10, $\rho=0.8$, $n=100$, $p=1000$. MSPEs maximum standard error is 0.02.} 
\centering
\resizebox{\textwidth}{!}{\begin{tabular}{lrrrrrrrrrrrr}
   \hline & \multicolumn{3}{c}{$\mathbf{\boldsymbol{\zeta}=0.05}$} & \multicolumn{3}{c}{$\mathbf{\boldsymbol{\zeta}=0.1}$} & \multicolumn{3}{c}{$\mathbf{\boldsymbol{\zeta}=0.2}$} & \multicolumn{3}{c}{$\mathbf{\boldsymbol{\zeta}=0.4}$} \\ \cmidrule(lr){2-4} \cmidrule(lr){5-7} \cmidrule(lr){8-10} \cmidrule(lr){11-13} \textbf{Method} & \textbf{MSPE} & \textbf{RC} & \textbf{PR} & \textbf{MSPE} & \textbf{RC} & \textbf{PR} & \textbf{MSPE} & \textbf{RC} & \textbf{PR} & \textbf{MSPE} & \textbf{RC} & \textbf{PR} \\ \cmidrule(lr){1-1} \cmidrule(lr){2-4} \cmidrule(lr){5-7} \cmidrule(lr){8-10} \cmidrule(lr){11-13}Lasso & 1.27 & 0.07 & 0.09 & 1.21 & 0.06 & 0.14 & 1.19 & 0.05 & 0.22 & 1.18 & 0.04 & 0.42 \\ 
  Elastic Net & 1.26 & 0.08 & 0.09 & 1.20 & 0.06 & 0.14 & 1.18 & 0.05 & 0.22 & 1.17 & 0.05 & 0.43 \\ 
  SplitReg-Lasso & 1.19 & 0.25 & 0.08 & 1.11 & 0.22 & 0.13 & 1.11 & 0.20 & 0.22 & 1.09 & 0.19 & 0.42 \\ 
  SplitReg-EN & 1.19 & 0.25 & 0.08 & 1.11 & 0.23 & 0.13 & 1.11 & 0.20 & 0.22 & 1.09 & 0.19 & 0.42 \\ 
  SparseNet & 1.31 & 0.09 & 0.12 & 1.23 & 0.07 & 0.18 & 1.22 & 0.06 & 0.28 & 1.22 & 0.05 & 0.48 \\ 
  Relaxed & 1.33 & 0.05 & 0.10 & 1.27 & 0.04 & 0.14 & 1.23 & 0.03 & 0.23 & 1.24 & 0.03 & 0.42 \\ 
  Stepwise & 1.74 & 0.01 & 0.11 & 1.70 & 0.01 & 0.14 & 1.65 & 0.01 & 0.22 & 1.60 & 0.01 & 0.41 \\ 
  RF & 1.32 & $ - $ & $ - $ & 1.25 & $ - $ & $ - $ & 1.25 & $ - $ & $ - $ & 1.25 & $ - $ & $ - $ \\ 
  RGLM & 1.21 & 0.73 & 0.05 & 1.13 & 0.70 & 0.10 & 1.12 & 0.69 & 0.20 & 1.10 & 0.68 & 0.40 \\ 
  SIS-SCAD & 2.25 & 0.03 & 0.07 & 2.14 & 0.02 & 0.11 & 2.08 & 0.02 & 0.21 & 2.05 & 0.02 & 0.40 \\ 
   \hline
\end{tabular}}
\end{table}
\begin{table}[ht]
\caption{Mean MSPEs, recalls and precisions for Scenario 1 with SNR = 3, $\rho=0.2$, $n=75$, $p=150$. MSPEs maximum standard error is 0.1.} 
\centering
\resizebox{\textwidth}{!}{\begin{tabular}{lrrrrrrrrrrrr}
   \hline & \multicolumn{3}{c}{$\mathbf{\boldsymbol{\zeta}=0.05}$} & \multicolumn{3}{c}{$\mathbf{\boldsymbol{\zeta}=0.1}$} & \multicolumn{3}{c}{$\mathbf{\boldsymbol{\zeta}=0.2}$} & \multicolumn{3}{c}{$\mathbf{\boldsymbol{\zeta}=0.4}$} \\ \cmidrule(lr){2-4} \cmidrule(lr){5-7} \cmidrule(lr){8-10} \cmidrule(lr){11-13} \textbf{Method} & \textbf{MSPE} & \textbf{RC} & \textbf{PR} & \textbf{MSPE} & \textbf{RC} & \textbf{PR} & \textbf{MSPE} & \textbf{RC} & \textbf{PR} & \textbf{MSPE} & \textbf{RC} & \textbf{PR} \\ \cmidrule(lr){1-1} \cmidrule(lr){2-4} \cmidrule(lr){5-7} \cmidrule(lr){8-10} \cmidrule(lr){11-13}Lasso & 1.79 & 0.98 & 0.26 & 2.01 & 0.63 & 0.34 & 1.98 & 0.42 & 0.39 & 1.72 & 0.31 & 0.52 \\ 
  Elastic Net & 1.83 & 0.97 & 0.24 & 1.99 & 0.65 & 0.33 & 1.94 & 0.43 & 0.39 & 1.71 & 0.33 & 0.51 \\ 
  SplitReg-Lasso & 1.78 & 0.98 & 0.27 & 1.99 & 0.69 & 0.30 & 1.85 & 0.62 & 0.32 & 1.55 & 0.68 & 0.46 \\ 
  SplitReg-EN & 1.82 & 0.98 & 0.24 & 1.99 & 0.69 & 0.30 & 1.85 & 0.61 & 0.33 & 1.53 & 0.68 & 0.46 \\ 
  SparseNet & 1.59 & 0.92 & 0.66 & 2.10 & 0.59 & 0.37 & 2.06 & 0.40 & 0.41 & 1.78 & 0.31 & 0.52 \\ 
  Relaxed & 1.78 & 0.90 & 0.54 & 2.09 & 0.53 & 0.42 & 2.05 & 0.34 & 0.44 & 1.83 & 0.25 & 0.53 \\ 
  Stepwise & 1.75 & 0.81 & 0.83 & 2.77 & 0.22 & 0.61 & 2.92 & 0.11 & 0.52 & 2.63 & 0.08 & 0.56 \\ 
  RF & $>3$  & $ - $ & $ - $ & 2.40 & $ - $ & $ - $ & 2.15 & $ - $ & $ - $ & 1.94 & $ - $ & $ - $ \\ 
  RGLM & $>3$  & 1.00 & 0.05 & 2.79 & 0.98 & 0.10 & $>3$  & 0.99 & 0.20 & $>3$  & 0.99 & 0.40 \\ 
  SIS-SCAD & 1.96 & 0.96 & 0.39 & 2.69 & 0.44 & 0.39 & 2.75 & 0.24 & 0.42 & 2.47 & 0.15 & 0.52 \\ 
  CRL & 1.90 & 0.95 & 0.26 & 1.98 & 0.64 & 0.33 & 1.91 & 0.42 & 0.37 & 1.70 & 0.32 & 0.52 \\ 
   \hline
\end{tabular}}
\end{table}
\begin{table}[ht]
\caption{Mean MSPEs, recalls and precisions for Scenario 1 with SNR = 5, $\rho=0.2$, $n=75$, $p=150$. MSPEs maximum standard error is 0.18.} 
\centering
\resizebox{\textwidth}{!}{\begin{tabular}{lrrrrrrrrrrrr}
   \hline & \multicolumn{3}{c}{$\mathbf{\boldsymbol{\zeta}=0.05}$} & \multicolumn{3}{c}{$\mathbf{\boldsymbol{\zeta}=0.1}$} & \multicolumn{3}{c}{$\mathbf{\boldsymbol{\zeta}=0.2}$} & \multicolumn{3}{c}{$\mathbf{\boldsymbol{\zeta}=0.4}$} \\ \cmidrule(lr){2-4} \cmidrule(lr){5-7} \cmidrule(lr){8-10} \cmidrule(lr){11-13} \textbf{Method} & \textbf{MSPE} & \textbf{RC} & \textbf{PR} & \textbf{MSPE} & \textbf{RC} & \textbf{PR} & \textbf{MSPE} & \textbf{RC} & \textbf{PR} & \textbf{MSPE} & \textbf{RC} & \textbf{PR} \\ \cmidrule(lr){1-1} \cmidrule(lr){2-4} \cmidrule(lr){5-7} \cmidrule(lr){8-10} \cmidrule(lr){11-13}Lasso & 1.79 & 0.99 & 0.25 & 2.29 & 0.75 & 0.35 & 2.38 & 0.50 & 0.41 & 1.99 & 0.37 & 0.53 \\ 
  Elastic Net & 1.86 & 1.00 & 0.23 & 2.28 & 0.76 & 0.34 & 2.35 & 0.51 & 0.41 & 1.97 & 0.38 & 0.53 \\ 
  SplitReg-Lasso & 1.79 & 1.00 & 0.26 & 2.31 & 0.78 & 0.33 & 2.25 & 0.67 & 0.34 & 1.71 & 0.76 & 0.45 \\ 
  SplitReg-EN & 1.85 & 1.00 & 0.23 & 2.31 & 0.78 & 0.32 & 2.24 & 0.66 & 0.34 & 1.70 & 0.76 & 0.45 \\ 
  SparseNet & 1.37 & 0.98 & 0.78 & 2.44 & 0.71 & 0.38 & 2.48 & 0.50 & 0.42 & 2.05 & 0.37 & 0.54 \\ 
  Relaxed & 1.64 & 0.97 & 0.60 & 2.35 & 0.66 & 0.44 & 2.49 & 0.41 & 0.46 & 2.09 & 0.30 & 0.55 \\ 
  Stepwise & 1.43 & 0.96 & 0.88 & $>3$  & 0.30 & 0.64 & $>3$  & 0.15 & 0.56 & $>3$  & 0.10 & 0.57 \\ 
  RF & $>3$  & $ - $ & $ - $ & $>3$  & $ - $ & $ - $ & 2.93 & $ - $ & $ - $ & 2.59 & $ - $ & $ - $ \\ 
  RGLM & $>3$  & 1.00 & 0.05 & $>3$  & 0.98 & 0.10 & $>3$  & 0.98 & 0.20 & 2.59 & 0.99 & 0.40 \\ 
  SIS-SCAD & 1.89 & 0.99 & 0.41 & $>3$  & 0.52 & 0.46 & $>3$  & 0.26 & 0.47 & 2.80 & 0.16 & 0.55 \\ 
  CRL & 2.00 & 0.98 & 0.24 & 2.28 & 0.76 & 0.34 & 2.32 & 0.50 & 0.39 & 1.97 & 0.38 & 0.53 \\ 
   \hline
\end{tabular}}
\end{table}
\begin{table}[ht]
\caption{Mean MSPEs, recalls and precisions for Scenario 1 with SNR = 10, $\rho=0.2$, $n=75$, $p=150$. MSPEs maximum standard error is 0.05.} 
\centering
\resizebox{\textwidth}{!}{\begin{tabular}{lrrrrrrrrrrrr}
   \hline & \multicolumn{3}{c}{$\mathbf{\boldsymbol{\zeta}=0.05}$} & \multicolumn{3}{c}{$\mathbf{\boldsymbol{\zeta}=0.1}$} & \multicolumn{3}{c}{$\mathbf{\boldsymbol{\zeta}=0.2}$} & \multicolumn{3}{c}{$\mathbf{\boldsymbol{\zeta}=0.4}$} \\ \cmidrule(lr){2-4} \cmidrule(lr){5-7} \cmidrule(lr){8-10} \cmidrule(lr){11-13} \textbf{Method} & \textbf{MSPE} & \textbf{RC} & \textbf{PR} & \textbf{MSPE} & \textbf{RC} & \textbf{PR} & \textbf{MSPE} & \textbf{RC} & \textbf{PR} & \textbf{MSPE} & \textbf{RC} & \textbf{PR} \\ \cmidrule(lr){1-1} \cmidrule(lr){2-4} \cmidrule(lr){5-7} \cmidrule(lr){8-10} \cmidrule(lr){11-13}Lasso & 1.79 & 1.00 & 0.25 & 2.70 & 0.89 & 0.35 & $>3$  & 0.61 & 0.43 & 2.55 & 0.45 & 0.55 \\ 
  Elastic Net & 1.88 & 1.00 & 0.23 & 2.71 & 0.89 & 0.33 & $>3$  & 0.62 & 0.42 & 2.52 & 0.46 & 0.55 \\ 
  SplitReg-Lasso & 1.79 & 1.00 & 0.26 & 2.78 & 0.89 & 0.34 & $>3$  & 0.73 & 0.36 & 2.10 & 0.82 & 0.45 \\ 
  SplitReg-EN & 1.88 & 1.00 & 0.23 & 2.77 & 0.90 & 0.33 & $>3$  & 0.74 & 0.35 & 2.09 & 0.82 & 0.45 \\ 
  SparseNet & 1.23 & 1.00 & 0.87 & 2.75 & 0.87 & 0.39 & $>3$  & 0.62 & 0.43 & 2.61 & 0.46 & 0.56 \\ 
  Relaxed & 1.46 & 1.00 & 0.70 & 2.80 & 0.81 & 0.45 & $>3$  & 0.52 & 0.48 & 2.68 & 0.38 & 0.58 \\ 
  Stepwise & 1.26 & 1.00 & 0.92 & $>3$  & 0.49 & 0.65 & $>3$  & 0.20 & 0.57 & $>3$  & 0.13 & 0.59 \\ 
  RF & $>3$  & $ - $ & $ - $ & $>3$  & $ - $ & $ - $ & $>3$  & $ - $ & $ - $ & $>3$  & $ - $ & $ - $ \\ 
  RGLM & $>3$  & 1.00 & 0.05 & $>3$  & 0.98 & 0.10 & $>3$  & 0.98 & 0.20 & $>3$  & 0.99 & 0.40 \\ 
  SIS-SCAD & 1.77 & 1.00 & 0.45 & $>3$  & 0.63 & 0.56 & $>3$  & 0.30 & 0.53 & $>3$  & 0.17 & 0.60 \\ 
  CRL & 2.33 & 0.99 & 0.24 & 2.72 & 0.89 & 0.34 & $>3$  & 0.61 & 0.41 & 2.53 & 0.45 & 0.56 \\ 
   \hline
\end{tabular}}
\end{table}
\begin{table}[ht]
\caption{Mean MSPEs, recalls and precisions for Scenario 1 with SNR = 3, $\rho=0.5$, $n=75$, $p=150$. MSPEs maximum standard error is 0.47.} 
\centering
\resizebox{\textwidth}{!}{\begin{tabular}{lrrrrrrrrrrrr}
   \hline & \multicolumn{3}{c}{$\mathbf{\boldsymbol{\zeta}=0.05}$} & \multicolumn{3}{c}{$\mathbf{\boldsymbol{\zeta}=0.1}$} & \multicolumn{3}{c}{$\mathbf{\boldsymbol{\zeta}=0.2}$} & \multicolumn{3}{c}{$\mathbf{\boldsymbol{\zeta}=0.4}$} \\ \cmidrule(lr){2-4} \cmidrule(lr){5-7} \cmidrule(lr){8-10} \cmidrule(lr){11-13} \textbf{Method} & \textbf{MSPE} & \textbf{RC} & \textbf{PR} & \textbf{MSPE} & \textbf{RC} & \textbf{PR} & \textbf{MSPE} & \textbf{RC} & \textbf{PR} & \textbf{MSPE} & \textbf{RC} & \textbf{PR} \\ \cmidrule(lr){1-1} \cmidrule(lr){2-4} \cmidrule(lr){5-7} \cmidrule(lr){8-10} \cmidrule(lr){11-13}Lasso & 1.81 & 0.97 & 0.25 & 1.57 & 0.43 & 0.28 & 1.45 & 0.27 & 0.34 & 1.36 & 0.21 & 0.48 \\ 
  Elastic Net & 1.84 & 0.97 & 0.23 & 1.55 & 0.45 & 0.28 & 1.43 & 0.29 & 0.33 & 1.35 & 0.22 & 0.48 \\ 
  SplitReg-Lasso & 1.79 & 0.97 & 0.26 & 1.50 & 0.59 & 0.22 & 1.36 & 0.55 & 0.28 & 1.24 & 0.56 & 0.44 \\ 
  SplitReg-EN & 1.84 & 0.97 & 0.24 & 1.50 & 0.60 & 0.22 & 1.35 & 0.57 & 0.27 & 1.24 & 0.55 & 0.45 \\ 
  SparseNet & 1.65 & 0.90 & 0.63 & 1.62 & 0.39 & 0.32 & 1.53 & 0.25 & 0.36 & 1.42 & 0.20 & 0.49 \\ 
  Relaxed & 1.78 & 0.90 & 0.51 & 1.63 & 0.35 & 0.33 & 1.53 & 0.22 & 0.38 & 1.44 & 0.15 & 0.49 \\ 
  Stepwise & 1.76 & 0.81 & 0.81 & 1.99 & 0.13 & 0.45 & 1.96 & 0.06 & 0.40 & 1.84 & 0.04 & 0.50 \\ 
  RF & $>3$  & $ - $ & $ - $ & 1.60 & $ - $ & $ - $ & 1.43 & $ - $ & $ - $ & 1.33 & $ - $ & $ - $ \\ 
  RGLM & $>3$  & 1.00 & 0.05 & 2.70 & 0.98 & 0.10 & 2.79 & 0.99 & 0.20 & $>3$  & 0.99 & 0.40 \\ 
  SIS-SCAD & 2.16 & 0.89 & 0.38 & 2.42 & 0.28 & 0.25 & 2.27 & 0.17 & 0.30 & 2.19 & 0.12 & 0.44 \\ 
  CRL & 1.80 & 0.95 & 0.25 & 1.51 & 0.41 & 0.29 & 1.40 & 0.29 & 0.35 & 1.30 & 0.22 & 0.49 \\ 
   \hline
\end{tabular}}
\end{table}
\begin{table}[ht]
\caption{Mean MSPEs, recalls and precisions for Scenario 1 with SNR = 5, $\rho=0.5$, $n=75$, $p=150$. MSPEs maximum standard error is 0.32.}
\centering
\resizebox{\textwidth}{!}{\begin{tabular}{lrrrrrrrrrrrr}
   \hline & \multicolumn{3}{c}{$\mathbf{\boldsymbol{\zeta}=0.05}$} & \multicolumn{3}{c}{$\mathbf{\boldsymbol{\zeta}=0.1}$} & \multicolumn{3}{c}{$\mathbf{\boldsymbol{\zeta}=0.2}$} & \multicolumn{3}{c}{$\mathbf{\boldsymbol{\zeta}=0.4}$} \\ \cmidrule(lr){2-4} \cmidrule(lr){5-7} \cmidrule(lr){8-10} \cmidrule(lr){11-13} \textbf{Method} & \textbf{MSPE} & \textbf{RC} & \textbf{PR} & \textbf{MSPE} & \textbf{RC} & \textbf{PR} & \textbf{MSPE} & \textbf{RC} & \textbf{PR} & \textbf{MSPE} & \textbf{RC} & \textbf{PR} \\ \cmidrule(lr){1-1} \cmidrule(lr){2-4} \cmidrule(lr){5-7} \cmidrule(lr){8-10} \cmidrule(lr){11-13}Lasso & 1.82 & 0.99 & 0.24 & 1.74 & 0.53 & 0.31 & 1.62 & 0.34 & 0.36 & 1.46 & 0.25 & 0.49 \\ 
  Elastic Net & 1.88 & 0.99 & 0.23 & 1.73 & 0.55 & 0.30 & 1.60 & 0.35 & 0.36 & 1.45 & 0.26 & 0.49 \\ 
  SplitReg-Lasso & 1.81 & 0.99 & 0.26 & 1.71 & 0.64 & 0.25 & 1.50 & 0.64 & 0.28 & 1.30 & 0.65 & 0.44 \\ 
  SplitReg-EN & 1.88 & 0.99 & 0.22 & 1.71 & 0.65 & 0.24 & 1.49 & 0.63 & 0.28 & 1.30 & 0.64 & 0.45 \\ 
  SparseNet & 1.38 & 0.98 & 0.75 & 1.82 & 0.50 & 0.34 & 1.70 & 0.33 & 0.38 & 1.50 & 0.25 & 0.51 \\ 
  Relaxed & 1.65 & 0.96 & 0.58 & 1.82 & 0.44 & 0.37 & 1.68 & 0.28 & 0.39 & 1.54 & 0.20 & 0.51 \\ 
  Stepwise & 1.48 & 0.95 & 0.85 & 2.34 & 0.18 & 0.49 & 2.31 & 0.09 & 0.43 & 2.07 & 0.06 & 0.51 \\ 
  RF & $>3$  & $ - $ & $ - $ & 1.98 & $ - $ & $ - $ & 1.72 & $ - $ & $ - $ & 1.56 & $ - $ & $ - $ \\ 
  RGLM & $>3$  & 1.00 & 0.05 & $>3$  & 0.98 & 0.10 & $>3$  & 0.99 & 0.20 & 2.49 & 0.99 & 0.40 \\ 
  SIS-SCAD & 2.12 & 0.94 & 0.43 & 2.62 & 0.35 & 0.31 & 2.47 & 0.19 & 0.34 & 2.25 & 0.13 & 0.46 \\ 
  CRL & 1.89 & 0.98 & 0.24 & 1.70 & 0.52 & 0.31 & 1.57 & 0.36 & 0.36 & 1.42 & 0.26 & 0.50 \\ 
   \hline
\end{tabular}} 
\end{table}
\begin{table}[ht]
\caption{Mean MSPEs, recalls and precisions for Scenario 1 with SNR = 10, $\rho=0.5$, $n=75$, $p=150$. MSPEs maximum standard error is 0.47.}
\centering
\resizebox{\textwidth}{!}{\begin{tabular}{lrrrrrrrrrrrr}
   \hline & \multicolumn{3}{c}{$\mathbf{\boldsymbol{\zeta}=0.05}$} & \multicolumn{3}{c}{$\mathbf{\boldsymbol{\zeta}=0.1}$} & \multicolumn{3}{c}{$\mathbf{\boldsymbol{\zeta}=0.2}$} & \multicolumn{3}{c}{$\mathbf{\boldsymbol{\zeta}=0.4}$} \\ \cmidrule(lr){2-4} \cmidrule(lr){5-7} \cmidrule(lr){8-10} \cmidrule(lr){11-13} \textbf{Method} & \textbf{MSPE} & \textbf{RC} & \textbf{PR} & \textbf{MSPE} & \textbf{RC} & \textbf{PR} & \textbf{MSPE} & \textbf{RC} & \textbf{PR} & \textbf{MSPE} & \textbf{RC} & \textbf{PR} \\ \cmidrule(lr){1-1} \cmidrule(lr){2-4} \cmidrule(lr){5-7} \cmidrule(lr){8-10} \cmidrule(lr){11-13}Lasso & 1.82 & 1.00 & 0.24 & 2.07 & 0.68 & 0.34 & 1.98 & 0.44 & 0.39 & 1.67 & 0.32 & 0.52 \\ 
  Elastic Net & 1.92 & 1.00 & 0.22 & 2.06 & 0.69 & 0.33 & 1.97 & 0.45 & 0.39 & 1.66 & 0.33 & 0.51 \\ 
  SplitReg-Lasso & 1.81 & 1.00 & 0.25 & 2.08 & 0.73 & 0.29 & 1.83 & 0.69 & 0.30 & 1.43 & 0.75 & 0.44 \\ 
  SplitReg-EN & 1.92 & 1.00 & 0.22 & 2.07 & 0.73 & 0.29 & 1.82 & 0.69 & 0.30 & 1.43 & 0.73 & 0.45 \\ 
  SparseNet & 1.25 & 1.00 & 0.85 & 2.18 & 0.65 & 0.37 & 2.05 & 0.44 & 0.41 & 1.70 & 0.33 & 0.53 \\ 
  Relaxed & 1.47 & 0.99 & 0.68 & 2.16 & 0.60 & 0.40 & 2.08 & 0.37 & 0.43 & 1.76 & 0.26 & 0.53 \\ 
  Stepwise & 1.29 & 1.00 & 0.90 & $>3$  & 0.27 & 0.53 & $>3$  & 0.12 & 0.46 & 2.52 & 0.09 & 0.54 \\ 
  RF & $>3$  & $ - $ & $ - $ & 2.96 & $ - $ & $ - $ & 2.48 & $ - $ & $ - $ & 2.15 & $ - $ & $ - $ \\ 
  RGLM & $>3$  & 1.00 & 0.05 & $>3$  & 0.98 & 0.10 & $>3$  & 0.99 & 0.20 & 2.91 & 0.99 & 0.40 \\ 
  SIS-SCAD & 1.93 & 0.98 & 0.53 & 2.98 & 0.46 & 0.40 & 2.95 & 0.24 & 0.42 & 2.51 & 0.14 & 0.50 \\ 
  CRL & 2.03 & 0.99 & 0.24 & 2.05 & 0.67 & 0.34 & 1.92 & 0.44 & 0.39 & 1.63 & 0.32 & 0.52 \\ 
   \hline
\end{tabular}} 
\end{table}
\begin{table}[ht]
\caption{Mean MSPEs, recalls and precisions for Scenario 1 with SNR = 3, $\rho=0.8$, $n=75$, $p=150$. MSPEs maximum standard error is 0.49.} 
\centering
\resizebox{\textwidth}{!}{\begin{tabular}{lrrrrrrrrrrrr}
   \hline & \multicolumn{3}{c}{$\mathbf{\boldsymbol{\zeta}=0.05}$} & \multicolumn{3}{c}{$\mathbf{\boldsymbol{\zeta}=0.1}$} & \multicolumn{3}{c}{$\mathbf{\boldsymbol{\zeta}=0.2}$} & \multicolumn{3}{c}{$\mathbf{\boldsymbol{\zeta}=0.4}$} \\ \cmidrule(lr){2-4} \cmidrule(lr){5-7} \cmidrule(lr){8-10} \cmidrule(lr){11-13} \textbf{Method} & \textbf{MSPE} & \textbf{RC} & \textbf{PR} & \textbf{MSPE} & \textbf{RC} & \textbf{PR} & \textbf{MSPE} & \textbf{RC} & \textbf{PR} & \textbf{MSPE} & \textbf{RC} & \textbf{PR} \\ \cmidrule(lr){1-1} \cmidrule(lr){2-4} \cmidrule(lr){5-7} \cmidrule(lr){8-10} \cmidrule(lr){11-13}Lasso & 1.81 & 0.96 & 0.25 & 1.27 & 0.23 & 0.21 & 1.22 & 0.16 & 0.28 & 1.20 & 0.13 & 0.44 \\ 
  Elastic Net & 1.86 & 0.96 & 0.23 & 1.27 & 0.25 & 0.21 & 1.20 & 0.17 & 0.28 & 1.19 & 0.14 & 0.44 \\ 
  SplitReg-Lasso & 1.79 & 0.96 & 0.26 & 1.22 & 0.45 & 0.17 & 1.16 & 0.43 & 0.25 & 1.14 & 0.39 & 0.43 \\ 
  SplitReg-EN & 1.87 & 0.96 & 0.22 & 1.22 & 0.47 & 0.16 & 1.16 & 0.44 & 0.25 & 1.14 & 0.39 & 0.43 \\ 
  SparseNet & 1.67 & 0.88 & 0.61 & 1.33 & 0.22 & 0.27 & 1.27 & 0.15 & 0.32 & 1.25 & 0.12 & 0.49 \\ 
  Relaxed & 1.78 & 0.88 & 0.49 & 1.34 & 0.18 & 0.26 & 1.26 & 0.11 & 0.29 & 1.26 & 0.10 & 0.48 \\ 
  Stepwise & 1.81 & 0.78 & 0.79 & 1.51 & 0.06 & 0.32 & 1.47 & 0.03 & 0.32 & 1.44 & 0.02 & 0.49 \\ 
  RF & $>3$  & $ - $ & $ - $ & 1.24 & $ - $ & $ - $ & 1.18 & $ - $ & $ - $ & 1.16 & $ - $ & $ - $ \\ 
  RGLM & $>3$  & 0.92 & 0.04 & $>3$  & 0.99 & 0.10 & 2.66 & 0.99 & 0.20 & $>3$  & 0.99 & 0.40 \\ 
  SIS-SCAD & 2.73 & 0.59 & 0.31 & 2.14 & 0.17 & 0.20 & 2.08 & 0.12 & 0.25 & 2.05 & 0.10 & 0.42 \\ 
  CRL & 1.75 & 0.93 & 0.26 & 1.24 & 0.23 & 0.22 & 1.18 & 0.17 & 0.28 & 1.14 & 0.13 & 0.46 \\ 
   \hline
\end{tabular}}
\end{table}
\begin{table}[ht]
\caption{Mean MSPEs, recalls and precisions for Scenario 1 with SNR = 5, $\rho=0.8$, $n=75$, $p=150$. MSPEs maximum standard error is 0.11.} 
\centering
\resizebox{\textwidth}{!}{\begin{tabular}{lrrrrrrrrrrrr}
   \hline & \multicolumn{3}{c}{$\mathbf{\boldsymbol{\zeta}=0.05}$} & \multicolumn{3}{c}{$\mathbf{\boldsymbol{\zeta}=0.1}$} & \multicolumn{3}{c}{$\mathbf{\boldsymbol{\zeta}=0.2}$} & \multicolumn{3}{c}{$\mathbf{\boldsymbol{\zeta}=0.4}$} \\ \cmidrule(lr){2-4} \cmidrule(lr){5-7} \cmidrule(lr){8-10} \cmidrule(lr){11-13} \textbf{Method} & \textbf{MSPE} & \textbf{RC} & \textbf{PR} & \textbf{MSPE} & \textbf{RC} & \textbf{PR} & \textbf{MSPE} & \textbf{RC} & \textbf{PR} & \textbf{MSPE} & \textbf{RC} & \textbf{PR} \\ \cmidrule(lr){1-1} \cmidrule(lr){2-4} \cmidrule(lr){5-7} \cmidrule(lr){8-10} \cmidrule(lr){11-13}Lasso & 1.81 & 0.99 & 0.24 & 1.36 & 0.30 & 0.23 & 1.28 & 0.20 & 0.30 & 1.24 & 0.16 & 0.45 \\ 
  Elastic Net & 1.92 & 0.99 & 0.22 & 1.35 & 0.32 & 0.23 & 1.26 & 0.21 & 0.30 & 1.23 & 0.17 & 0.45 \\ 
  SplitReg-Lasso & 1.81 & 0.99 & 0.25 & 1.30 & 0.53 & 0.18 & 1.20 & 0.50 & 0.26 & 1.15 & 0.46 & 0.43 \\ 
  SplitReg-EN & 1.92 & 0.99 & 0.21 & 1.30 & 0.54 & 0.18 & 1.20 & 0.51 & 0.25 & 1.16 & 0.48 & 0.43 \\ 
  SparseNet & 1.41 & 0.97 & 0.74 & 1.42 & 0.28 & 0.29 & 1.35 & 0.19 & 0.34 & 1.29 & 0.15 & 0.50 \\ 
  Relaxed & 1.69 & 0.95 & 0.56 & 1.44 & 0.23 & 0.29 & 1.34 & 0.15 & 0.32 & 1.29 & 0.12 & 0.47 \\ 
  Stepwise & 1.60 & 0.92 & 0.82 & 1.67 & 0.08 & 0.35 & 1.60 & 0.04 & 0.35 & 1.54 & 0.03 & 0.49 \\ 
  RF & $>3$  & $ - $ & $ - $ & 1.36 & $ - $ & $ - $ & 1.27 & $ - $ & $ - $ & 1.22 & $ - $ & $ - $ \\ 
  RGLM & $>3$  & 0.91 & 0.04 & $>3$  & 0.98 & 0.10 & $>3$  & 0.99 & 0.20 & 2.14 & 0.99 & 0.40 \\ 
  SIS-SCAD & 2.97 & 0.72 & 0.37 & 2.27 & 0.20 & 0.21 & 2.15 & 0.13 & 0.26 & 2.14 & 0.11 & 0.42 \\ 
  CRL & 1.81 & 0.98 & 0.25 & 1.32 & 0.29 & 0.24 & 1.24 & 0.20 & 0.30 & 1.18 & 0.16 & 0.47 \\ 
   \hline
\end{tabular}}
\end{table}
\begin{table}[ht]
\caption{Mean MSPEs, recalls and precisions for Scenario 1 with SNR = 10, $\rho=0.8$, $n=75$, $p=150$. MSPEs maximum standard error is 0.55.} 
\centering
\resizebox{\textwidth}{!}{\begin{tabular}{lrrrrrrrrrrrr}
   \hline & \multicolumn{3}{c}{$\mathbf{\boldsymbol{\zeta}=0.05}$} & \multicolumn{3}{c}{$\mathbf{\boldsymbol{\zeta}=0.1}$} & \multicolumn{3}{c}{$\mathbf{\boldsymbol{\zeta}=0.2}$} & \multicolumn{3}{c}{$\mathbf{\boldsymbol{\zeta}=0.4}$} \\ \cmidrule(lr){2-4} \cmidrule(lr){5-7} \cmidrule(lr){8-10} \cmidrule(lr){11-13} \textbf{Method} & \textbf{MSPE} & \textbf{RC} & \textbf{PR} & \textbf{MSPE} & \textbf{RC} & \textbf{PR} & \textbf{MSPE} & \textbf{RC} & \textbf{PR} & \textbf{MSPE} & \textbf{RC} & \textbf{PR} \\ \cmidrule(lr){1-1} \cmidrule(lr){2-4} \cmidrule(lr){5-7} \cmidrule(lr){8-10} \cmidrule(lr){11-13}Lasso & 1.81 & 1.00 & 0.24 & 1.52 & 0.42 & 0.28 & 1.40 & 0.26 & 0.33 & 1.31 & 0.20 & 0.47 \\ 
  Elastic Net & 1.97 & 1.00 & 0.21 & 1.50 & 0.43 & 0.27 & 1.39 & 0.28 & 0.33 & 1.31 & 0.21 & 0.47 \\ 
  SplitReg-Lasso & 1.81 & 1.00 & 0.25 & 1.46 & 0.60 & 0.21 & 1.30 & 0.60 & 0.26 & 1.19 & 0.59 & 0.44 \\ 
  SplitReg-EN & 1.98 & 1.00 & 0.21 & 1.46 & 0.62 & 0.21 & 1.30 & 0.61 & 0.26 & 1.20 & 0.57 & 0.44 \\ 
  SparseNet & 1.27 & 1.00 & 0.84 & 1.59 & 0.38 & 0.33 & 1.47 & 0.26 & 0.36 & 1.35 & 0.21 & 0.51 \\ 
  Relaxed & 1.52 & 0.99 & 0.64 & 1.59 & 0.34 & 0.33 & 1.48 & 0.20 & 0.36 & 1.38 & 0.15 & 0.49 \\ 
  Stepwise & 1.33 & 1.00 & 0.87 & 1.98 & 0.13 & 0.40 & 1.87 & 0.07 & 0.40 & 1.75 & 0.04 & 0.48 \\ 
  RF & $>3$  & $ - $ & $ - $ & 1.67 & $ - $ & $ - $ & 1.49 & $ - $ & $ - $ & 1.38 & $ - $ & $ - $ \\ 
  RGLM & $>3$  & 0.90 & 0.04 & $>3$  & 0.98 & 0.10 & 2.33 & 0.99 & 0.20 & 2.80 & 0.99 & 0.40 \\ 
  SIS-SCAD & $>3$  & 0.83 & 0.45 & 2.51 & 0.28 & 0.25 & 2.34 & 0.16 & 0.30 & 2.23 & 0.12 & 0.43 \\ 
  CRL & 1.91 & 0.99 & 0.24 & 1.48 & 0.41 & 0.28 & 1.36 & 0.27 & 0.34 & 1.25 & 0.21 & 0.49 \\ 
   \hline
\end{tabular}}
\end{table}
\begin{table}[ht]
\caption{Mean MSPEs, recalls and precisions for Scenario 2 with SNR = 3, $\rho=0.2$, $n=100$, $p=1000$. MSPEs maximum standard error is 0.03.} 
\centering
\resizebox{\textwidth}{!}{\begin{tabular}{lrrrrrrrrrrrr}
   \hline & \multicolumn{3}{c}{$\mathbf{\boldsymbol{\zeta}=0.05}$} & \multicolumn{3}{c}{$\mathbf{\boldsymbol{\zeta}=0.1}$} & \multicolumn{3}{c}{$\mathbf{\boldsymbol{\zeta}=0.2}$} & \multicolumn{3}{c}{$\mathbf{\boldsymbol{\zeta}=0.4}$} \\ \cmidrule(lr){2-4} \cmidrule(lr){5-7} \cmidrule(lr){8-10} \cmidrule(lr){11-13} \textbf{Method} & \textbf{MSPE} & \textbf{RC} & \textbf{PR} & \textbf{MSPE} & \textbf{RC} & \textbf{PR} & \textbf{MSPE} & \textbf{RC} & \textbf{PR} & \textbf{MSPE} & \textbf{RC} & \textbf{PR} \\ \cmidrule(lr){1-1} \cmidrule(lr){2-4} \cmidrule(lr){5-7} \cmidrule(lr){8-10} \cmidrule(lr){11-13}Lasso & 2.14 & 0.18 & 0.15 & 1.91 & 0.12 & 0.18 & 1.76 & 0.10 & 0.26 & 1.64 & 0.09 & 0.44 \\ 
  Elastic Net & 2.12 & 0.19 & 0.15 & 1.88 & 0.13 & 0.18 & 1.74 & 0.11 & 0.26 & 1.63 & 0.09 & 0.44 \\ 
  SplitReg-Lasso & 1.92 & 0.48 & 0.10 & 1.63 & 0.44 & 0.14 & 1.48 & 0.43 & 0.23 & 1.33 & 0.46 & 0.42 \\ 
  SplitReg-EN & 1.93 & 0.47 & 0.10 & 1.62 & 0.45 & 0.14 & 1.47 & 0.44 & 0.23 & 1.33 & 0.46 & 0.42 \\ 
  SparseNet & 2.17 & 0.18 & 0.15 & 1.93 & 0.14 & 0.19 & 1.80 & 0.11 & 0.27 & 1.66 & 0.10 & 0.46 \\ 
  Relaxed & 2.26 & 0.14 & 0.17 & 1.99 & 0.10 & 0.18 & 1.84 & 0.08 & 0.27 & 1.72 & 0.07 & 0.45 \\ 
  Stepwise &$>3$& 0.02 & 0.20 &$>3$& 0.01 & 0.18 &$>3$& 0.01 & 0.26 &$>3$& 0.01 & 0.45 \\ 
  RF & 2.44 & $ - $ & $ - $ & 2.29 & $ - $ & $ - $ & 2.24 & $ - $ & $ - $ & 2.18 & $ - $ & $ - $ \\ 
  RGLM & 1.82 & 0.75 & 0.06 & 1.56 & 0.75 & 0.11 & 1.45 & 0.72 & 0.21 & 1.34 & 0.71 & 0.41 \\ 
  SIS-SCAD &$>3$& 0.07 & 0.16 & 2.84 & 0.04 & 0.17 & 2.64 & 0.03 & 0.27 & 2.48 & 0.02 & 0.44 \\ 
   \hline
\end{tabular}}
\end{table}
\begin{table}[ht]
\caption{Mean MSPEs, recalls and precisions for Scenario 2 with SNR = 5, $\rho=0.2$, $n=100$, $p=1000$. MSPEs maximum standard error is 0.03.}
\centering
\resizebox{\textwidth}{!}{\begin{tabular}{lrrrrrrrrrrrr}
   \hline & \multicolumn{3}{c}{$\mathbf{\boldsymbol{\zeta}=0.05}$} & \multicolumn{3}{c}{$\mathbf{\boldsymbol{\zeta}=0.1}$} & \multicolumn{3}{c}{$\mathbf{\boldsymbol{\zeta}=0.2}$} & \multicolumn{3}{c}{$\mathbf{\boldsymbol{\zeta}=0.4}$} \\ \cmidrule(lr){2-4} \cmidrule(lr){5-7} \cmidrule(lr){8-10} \cmidrule(lr){11-13} \textbf{Method} & \textbf{MSPE} & \textbf{RC} & \textbf{PR} & \textbf{MSPE} & \textbf{RC} & \textbf{PR} & \textbf{MSPE} & \textbf{RC} & \textbf{PR} & \textbf{MSPE} & \textbf{RC} & \textbf{PR} \\ \cmidrule(lr){1-1} \cmidrule(lr){2-4} \cmidrule(lr){5-7} \cmidrule(lr){8-10} \cmidrule(lr){11-13}Lasso & 2.70 & 0.23 & 0.17 & 2.35 & 0.15 & 0.19 & 2.08 & 0.12 & 0.27 & 1.90 & 0.10 & 0.45 \\ 
  Elastic Net & 2.67 & 0.24 & 0.17 & 2.32 & 0.16 & 0.19 & 2.06 & 0.13 & 0.28 & 1.88 & 0.11 & 0.45 \\ 
  SplitReg-Lasso & 2.40 & 0.55 & 0.10 & 1.95 & 0.54 & 0.14 & 1.67 & 0.52 & 0.23 & 1.44 & 0.56 & 0.42 \\ 
  SplitReg-EN & 2.40 & 0.55 & 0.11 & 1.94 & 0.54 & 0.14 & 1.65 & 0.54 & 0.23 & 1.43 & 0.57 & 0.42 \\ 
  SparseNet & 2.74 & 0.24 & 0.17 & 2.38 & 0.17 & 0.20 & 2.08 & 0.14 & 0.29 & 1.91 & 0.12 & 0.47 \\ 
  Relaxed & 2.83 & 0.19 & 0.19 & 2.46 & 0.13 & 0.20 & 2.18 & 0.10 & 0.28 & 1.97 & 0.09 & 0.46 \\ 
  Stepwise &$>3$& 0.03 & 0.20 &$>3$& 0.02 & 0.19 &$>3$& 0.02 & 0.26 &$>3$& 0.02 & 0.46 \\ 
  RF &$>3$& $ - $ & $ - $ &$>3$& $ - $ & $ - $ &$>3$& $ - $ & $ - $ &$>3$& $ - $ & $ - $ \\ 
  RGLM & 2.36 & 0.75 & 0.06 & 1.97 & 0.76 & 0.11 & 1.78 & 0.73 & 0.21 & 1.63 & 0.72 & 0.41 \\ 
  SIS-SCAD &$>3$& 0.07 & 0.18 &$>3$& 0.04 & 0.19 &$>3$& 0.03 & 0.27 &$>3$& 0.02 & 0.46 \\ 
   \hline
\end{tabular}} 
\end{table}
\begin{table}[ht]
\caption{Mean MSPEs, recalls and precisions for Scenario 2 with SNR = 10, $\rho=0.2$, $n=100$, $p=1000$. MSPEs maximum standard error is 0.03.} 
\centering
\resizebox{\textwidth}{!}{\begin{tabular}{lrrrrrrrrrrrr}
   \hline & \multicolumn{3}{c}{$\mathbf{\boldsymbol{\zeta}=0.05}$} & \multicolumn{3}{c}{$\mathbf{\boldsymbol{\zeta}=0.1}$} & \multicolumn{3}{c}{$\mathbf{\boldsymbol{\zeta}=0.2}$} & \multicolumn{3}{c}{$\mathbf{\boldsymbol{\zeta}=0.4}$} \\ \cmidrule(lr){2-4} \cmidrule(lr){5-7} \cmidrule(lr){8-10} \cmidrule(lr){11-13} \textbf{Method} & \textbf{MSPE} & \textbf{RC} & \textbf{PR} & \textbf{MSPE} & \textbf{RC} & \textbf{PR} & \textbf{MSPE} & \textbf{RC} & \textbf{PR} & \textbf{MSPE} & \textbf{RC} & \textbf{PR} \\ \cmidrule(lr){1-1} \cmidrule(lr){2-4} \cmidrule(lr){5-7} \cmidrule(lr){8-10} \cmidrule(lr){11-13}Lasso &$>3$& 0.29 & 0.19 &$>3$& 0.19 & 0.21 & 2.95 & 0.14 & 0.29 & 2.68 & 0.11 & 0.46 \\ 
  Elastic Net &$>3$& 0.30 & 0.19 &$>3$& 0.20 & 0.21 & 2.90 & 0.14 & 0.29 & 2.62 & 0.12 & 0.46 \\ 
  SplitReg-Lasso &$>3$& 0.61 & 0.11 & 2.71 & 0.63 & 0.14 & 2.10 & 0.66 & 0.23 & 1.71 & 0.69 & 0.42 \\ 
  SplitReg-EN &$>3$& 0.61 & 0.11 & 2.70 & 0.63 & 0.14 & 2.09 & 0.65 & 0.23 & 1.70 & 0.69 & 0.42 \\ 
  SparseNet &$>3$& 0.32 & 0.19 &$>3$& 0.23 & 0.22 & 2.88 & 0.17 & 0.31 & 2.59 & 0.14 & 0.49 \\ 
  Relaxed &$>3$& 0.24 & 0.21 &$>3$& 0.17 & 0.22 &$>3$& 0.12 & 0.29 & 2.76 & 0.10 & 0.47 \\ 
  Stepwise &$>3$& 0.04 & 0.22 &$>3$& 0.03 & 0.21 &$>3$& 0.02 & 0.27 &$>3$& 0.02 & 0.45 \\ 
  RF &$>3$& $ - $ & $ - $ &$>3$& $ - $ & $ - $ &$>3$& $ - $ & $ - $ &$>3$& $ - $ & $ - $ \\ 
  RGLM &$>3$& 0.76 & 0.06 &$>3$& 0.77 & 0.11 & 2.66 & 0.74 & 0.21 & 2.39 & 0.72 & 0.41 \\ 
  SIS-SCAD &$>3$& 0.09 & 0.21 &$>3$& 0.04 & 0.21 &$>3$& 0.03 & 0.29 &$>3$& 0.02 & 0.46 \\ 
   \hline
\end{tabular}}
\end{table}
\begin{table}[ht]
\caption{Mean MSPEs, recalls and precisions for Scenario 2 with SNR = 3, $\rho=0.5$, $n=100$, $p=1000$. MSPEs maximum standard error is 0.03.} 
\centering
\resizebox{\textwidth}{!}{\begin{tabular}{lrrrrrrrrrrrr}
   \hline & \multicolumn{3}{c}{$\mathbf{\boldsymbol{\zeta}=0.05}$} & \multicolumn{3}{c}{$\mathbf{\boldsymbol{\zeta}=0.1}$} & \multicolumn{3}{c}{$\mathbf{\boldsymbol{\zeta}=0.2}$} & \multicolumn{3}{c}{$\mathbf{\boldsymbol{\zeta}=0.4}$} \\ \cmidrule(lr){2-4} \cmidrule(lr){5-7} \cmidrule(lr){8-10} \cmidrule(lr){11-13} \textbf{Method} & \textbf{MSPE} & \textbf{RC} & \textbf{PR} & \textbf{MSPE} & \textbf{RC} & \textbf{PR} & \textbf{MSPE} & \textbf{RC} & \textbf{PR} & \textbf{MSPE} & \textbf{RC} & \textbf{PR} \\ \cmidrule(lr){1-1} \cmidrule(lr){2-4} \cmidrule(lr){5-7} \cmidrule(lr){8-10} \cmidrule(lr){11-13}Lasso & 1.48 & 0.10 & 0.11 & 1.37 & 0.08 & 0.15 & 1.33 & 0.06 & 0.23 & 1.29 & 0.06 & 0.42 \\ 
  Elastic Net & 1.47 & 0.11 & 0.11 & 1.35 & 0.08 & 0.15 & 1.33 & 0.07 & 0.24 & 1.28 & 0.06 & 0.42 \\ 
  SplitReg-Lasso & 1.35 & 0.33 & 0.08 & 1.23 & 0.29 & 0.13 & 1.21 & 0.27 & 0.23 & 1.16 & 0.25 & 0.42 \\ 
  SplitReg-EN & 1.36 & 0.33 & 0.09 & 1.23 & 0.29 & 0.13 & 1.20 & 0.28 & 0.22 & 1.16 & 0.26 & 0.42 \\ 
  SparseNet & 1.51 & 0.11 & 0.12 & 1.39 & 0.08 & 0.16 & 1.36 & 0.07 & 0.25 & 1.33 & 0.06 & 0.44 \\ 
  Relaxed & 1.53 & 0.08 & 0.12 & 1.42 & 0.06 & 0.15 & 1.38 & 0.05 & 0.24 & 1.35 & 0.04 & 0.43 \\ 
  Stepwise & 2.18 & 0.01 & 0.12 & 2.04 & 0.01 & 0.15 & 1.99 & 0.01 & 0.25 & 1.97 & 0.01 & 0.44 \\ 
  RF & 1.63 & $ - $ & $ - $ & 1.56 & $ - $ & $ - $ & 1.54 & $ - $ & $ - $ & 1.52 & $ - $ & $ - $ \\ 
  RGLM & 1.52 & 0.63 & 0.05 & 1.24 & 0.72 & 0.11 & 1.22 & 0.70 & 0.20 & 1.17 & 0.70 & 0.40 \\ 
  SIS-SCAD & 2.39 & 0.04 & 0.09 & 2.24 & 0.03 & 0.12 & 2.15 & 0.02 & 0.23 & 2.09 & 0.02 & 0.42 \\ 
   \hline
\end{tabular}}
\end{table}
\begin{table}[ht]
\caption{Mean MSPEs, recalls and precisions for Scenario 2 with SNR = 5, $\rho=0.5$, $n=100$, $p=1000$. MSPEs maximum standard error is 0.03.} 
\centering
\resizebox{\textwidth}{!}{\begin{tabular}{lrrrrrrrrrrrr}
   \hline & \multicolumn{3}{c}{$\mathbf{\boldsymbol{\zeta}=0.05}$} & \multicolumn{3}{c}{$\mathbf{\boldsymbol{\zeta}=0.1}$} & \multicolumn{3}{c}{$\mathbf{\boldsymbol{\zeta}=0.2}$} & \multicolumn{3}{c}{$\mathbf{\boldsymbol{\zeta}=0.4}$} \\ \cmidrule(lr){2-4} \cmidrule(lr){5-7} \cmidrule(lr){8-10} \cmidrule(lr){11-13} \textbf{Method} & \textbf{MSPE} & \textbf{RC} & \textbf{PR} & \textbf{MSPE} & \textbf{RC} & \textbf{PR} & \textbf{MSPE} & \textbf{RC} & \textbf{PR} & \textbf{MSPE} & \textbf{RC} & \textbf{PR} \\ \cmidrule(lr){1-1} \cmidrule(lr){2-4} \cmidrule(lr){5-7} \cmidrule(lr){8-10} \cmidrule(lr){11-13}Lasso & 1.69 & 0.14 & 0.12 & 1.52 & 0.10 & 0.16 & 1.45 & 0.08 & 0.24 & 1.39 & 0.07 & 0.43 \\ 
  Elastic Net & 1.67 & 0.15 & 0.12 & 1.50 & 0.10 & 0.16 & 1.44 & 0.08 & 0.25 & 1.37 & 0.07 & 0.43 \\ 
  SplitReg-Lasso & 1.51 & 0.41 & 0.09 & 1.33 & 0.37 & 0.13 & 1.26 & 0.34 & 0.23 & 1.20 & 0.32 & 0.42 \\ 
  SplitReg-EN & 1.51 & 0.41 & 0.09 & 1.32 & 0.37 & 0.13 & 1.25 & 0.35 & 0.23 & 1.20 & 0.33 & 0.42 \\ 
  SparseNet & 1.71 & 0.15 & 0.14 & 1.54 & 0.11 & 0.17 & 1.47 & 0.09 & 0.27 & 1.41 & 0.08 & 0.45 \\ 
  Relaxed & 1.76 & 0.11 & 0.13 & 1.59 & 0.07 & 0.16 & 1.53 & 0.06 & 0.25 & 1.45 & 0.05 & 0.43 \\ 
  Stepwise & 2.62 & 0.02 & 0.13 & 2.46 & 0.01 & 0.15 & 2.36 & 0.01 & 0.25 & 2.31 & 0.01 & 0.44 \\ 
  RF & 2.09 & $ - $ & $ - $ & 2.00 & $ - $ & $ - $ & 1.96 & $ - $ & $ - $ & 1.93 & $ - $ & $ - $ \\ 
  RGLM & 1.90 & 0.62 & 0.05 & 1.38 & 0.73 & 0.11 & 1.34 & 0.71 & 0.21 & 1.28 & 0.70 & 0.40 \\ 
  SIS-SCAD & 2.73 & 0.05 & 0.11 & 2.50 & 0.03 & 0.14 & 2.33 & 0.03 & 0.24 & 2.24 & 0.02 & 0.43 \\ 
   \hline
\end{tabular}}
\end{table}
\begin{table}[ht]
\caption{Mean MSPEs, recalls and precisions for Scenario 2 with SNR = 10, $\rho=0.5$, $n=100$, $p=1000$. MSPEs maximum standard error is 0.05.} 
\centering
\resizebox{\textwidth}{!}{\begin{tabular}{lrrrrrrrrrrrr}
   \hline & \multicolumn{3}{c}{$\mathbf{\boldsymbol{\zeta}=0.05}$} & \multicolumn{3}{c}{$\mathbf{\boldsymbol{\zeta}=0.1}$} & \multicolumn{3}{c}{$\mathbf{\boldsymbol{\zeta}=0.2}$} & \multicolumn{3}{c}{$\mathbf{\boldsymbol{\zeta}=0.4}$} \\ \cmidrule(lr){2-4} \cmidrule(lr){5-7} \cmidrule(lr){8-10} \cmidrule(lr){11-13} \textbf{Method} & \textbf{MSPE} & \textbf{RC} & \textbf{PR} & \textbf{MSPE} & \textbf{RC} & \textbf{PR} & \textbf{MSPE} & \textbf{RC} & \textbf{PR} & \textbf{MSPE} & \textbf{RC} & \textbf{PR} \\ \cmidrule(lr){1-1} \cmidrule(lr){2-4} \cmidrule(lr){5-7} \cmidrule(lr){8-10} \cmidrule(lr){11-13}Lasso & 2.15 & 0.20 & 0.15 & 1.87 & 0.13 & 0.18 & 1.70 & 0.10 & 0.26 & 1.59 & 0.09 & 0.44 \\ 
  Elastic Net & 2.12 & 0.20 & 0.15 & 1.84 & 0.13 & 0.18 & 1.68 & 0.11 & 0.26 & 1.57 & 0.09 & 0.44 \\ 
  SplitReg-Lasso & 1.89 & 0.51 & 0.10 & 1.56 & 0.47 & 0.14 & 1.39 & 0.44 & 0.23 & 1.27 & 0.45 & 0.42 \\ 
  SplitReg-EN & 1.89 & 0.51 & 0.10 & 1.54 & 0.48 & 0.14 & 1.38 & 0.45 & 0.23 & 1.26 & 0.47 & 0.42 \\ 
  SparseNet & 2.15 & 0.23 & 0.16 & 1.86 & 0.16 & 0.20 & 1.68 & 0.13 & 0.29 & 1.60 & 0.11 & 0.48 \\ 
  Relaxed & 2.26 & 0.16 & 0.16 & 1.97 & 0.10 & 0.18 & 1.77 & 0.08 & 0.27 & 1.65 & 0.07 & 0.44 \\ 
  Stepwise &$>3$& 0.03 & 0.17 &$>3$& 0.02 & 0.17 &$>3$& 0.02 & 0.26 & 2.98 & 0.01 & 0.44 \\ 
  RF &$>3$& $ - $ & $ - $ &$>3$& $ - $ & $ - $ &$>3$& $ - $ & $ - $ & 2.98 & $ - $ & $ - $ \\ 
  RGLM & 2.90 & 0.62 & 0.05 & 1.80 & 0.74 & 0.11 & 1.70 & 0.72 & 0.21 & 1.58 & 0.71 & 0.40 \\ 
  SIS-SCAD &$>3$& 0.06 & 0.14 &$>3$& 0.04 & 0.17 & 2.77 & 0.03 & 0.25 & 2.63 & 0.02 & 0.45 \\ 
   \hline
\end{tabular}}
\end{table}
\begin{table}[ht]
\caption{Mean MSPEs, recalls and precisions for Scenario 2 with SNR = 3, $\rho=0.8$, $n=100$, $p=1000$. MSPEs maximum standard error is 0.04.} 
\centering
\resizebox{\textwidth}{!}{\begin{tabular}{lrrrrrrrrrrrr}
   \hline & \multicolumn{3}{c}{$\mathbf{\boldsymbol{\zeta}=0.05}$} & \multicolumn{3}{c}{$\mathbf{\boldsymbol{\zeta}=0.1}$} & \multicolumn{3}{c}{$\mathbf{\boldsymbol{\zeta}=0.2}$} & \multicolumn{3}{c}{$\mathbf{\boldsymbol{\zeta}=0.4}$} \\ \cmidrule(lr){2-4} \cmidrule(lr){5-7} \cmidrule(lr){8-10} \cmidrule(lr){11-13} \textbf{Method} & \textbf{MSPE} & \textbf{RC} & \textbf{PR} & \textbf{MSPE} & \textbf{RC} & \textbf{PR} & \textbf{MSPE} & \textbf{RC} & \textbf{PR} & \textbf{MSPE} & \textbf{RC} & \textbf{PR} \\ \cmidrule(lr){1-1} \cmidrule(lr){2-4} \cmidrule(lr){5-7} \cmidrule(lr){8-10} \cmidrule(lr){11-13}Lasso & 1.22 & 0.05 & 0.08 & 1.16 & 0.04 & 0.12 & 1.17 & 0.04 & 0.22 & 1.15 & 0.03 & 0.41 \\ 
  Elastic Net & 1.21 & 0.06 & 0.08 & 1.15 & 0.04 & 0.12 & 1.16 & 0.04 & 0.22 & 1.14 & 0.04 & 0.41 \\ 
  SplitReg-Lasso & 1.16 & 0.19 & 0.07 & 1.10 & 0.17 & 0.12 & 1.11 & 0.16 & 0.21 & 1.10 & 0.14 & 0.41 \\ 
  SplitReg-EN & 1.16 & 0.20 & 0.07 & 1.10 & 0.16 & 0.12 & 1.11 & 0.17 & 0.22 & 1.09 & 0.15 & 0.41 \\ 
  SparseNet & 1.26 & 0.06 & 0.10 & 1.21 & 0.04 & 0.14 & 1.20 & 0.04 & 0.25 & 1.21 & 0.04 & 0.44 \\ 
  Relaxed & 1.25 & 0.04 & 0.08 & 1.21 & 0.03 & 0.12 & 1.22 & 0.03 & 0.22 & 1.20 & 0.03 & 0.41 \\ 
  Stepwise & 1.51 & 0.01 & 0.08 & 1.46 & $ 0.00 $ & 0.12 & 1.46 & $ 0.00 $ & 0.21 & 1.45 & $ 0.00 $ & 0.41 \\ 
  RF & 1.34 & $ - $ & $ - $ & 1.29 & $ - $ & $ - $ & 1.31 & $ - $ & $ - $ & 1.30 & $ - $ & $ - $ \\ 
  RGLM & 1.74 & 0.52 & 0.05 & 1.27 & 0.69 & 0.10 & 1.29 & 0.68 & 0.20 & 1.28 & 0.67 & 0.40 \\ 
  SIS-SCAD & 2.50 & 0.03 & 0.06 & 2.19 & 0.02 & 0.11 & 2.23 & 0.02 & 0.20 & 2.20 & 0.02 & 0.41 \\ 
   \hline
\end{tabular}%
} 
\end{table}
\begin{table}[ht]
\caption{Mean MSPEs, recalls and precisions for Scenario 3 with SNR = 3, $\rho=0.2$, $n=100$, $p=1000$. MSPEs maximum standard error is 0.03.} 
\centering
\resizebox{\textwidth}{!}{%
}
\end{table}
\begin{table}[ht]
\caption{Mean MSPEs, recalls and precisions for Scenario 3 with SNR = 5, $\rho=0.2$, $n=100$, $p=1000$. MSPEs maximum standard error is 0.04.} 
\centering
\resizebox{\textwidth}{!}{%
}
\end{table}
\begin{table}[ht]
\caption{Mean MSPEs, recalls and precisions for Scenario 3 with SNR = 10, $\rho=0.2$, $n=100$, $p=1000$. MSPEs maximum standard error is 0.07.} 
\centering
\resizebox{\textwidth}{!}{%
}
\end{table}